\documentclass[10pt,journal]{IEEEtran}
\usepackage{cite}
\usepackage{amsmath}
\usepackage{amssymb}
\usepackage{floatfig,epsfig}
\usepackage{subfigure}
\usepackage{verbatim}
\usepackage{pifont}
\usepackage{algorithm}
\usepackage{algorithmic}
\usepackage{alltt}
\usepackage{bm}
\usepackage{isolatin1}
\usepackage{graphics}
\usepackage{graphicx,dblfloatfix}
\usepackage{epic}
\usepackage{eepic}
\usepackage{graphpap}
\usepackage{psfrag}
\usepackage{threeparttable}
\usepackage{arydshln}
\usepackage[usenames,dvipsnames]{pstricks}
\usepackage{epsfig}
\usepackage{pst-grad} % For gradients
\usepackage{pst-plot} % For axes
\newtheorem{definition}{\textbf{Definition}}
\newtheorem{lemma}{\textbf{Lemma}}
\newtheorem{theorem}{\textbf{Theorem}}

\newtheorem{remark}{\textbf{Remark}}

\normalsize

\ifCLASSINFOpdf
\else
\fi

\begin{document}
\title{Decentralized Caching Schemes and Performance Limits in Two-layer Networks}
\author{\authorblockN{Lin Zhang, Zhao Wang, Ming Xiao, \IEEEmembership{Senior Member, IEEE},\\ Gang Wu, \IEEEmembership{Member, IEEE}, Ying-Chang Liang, \IEEEmembership{Fellow, IEEE}, \\and Shaoqian Li, \IEEEmembership{Fellow, IEEE}}

\thanks{Copyright (c) 2015 IEEE. Personal use of this material is permitted. However, permission to use this material for any other purposes must be obtained from the IEEE by sending a request to pubs-permissions@ieee.org. }
\thanks{ L. Zhang is with the Key Laboratory on Communications, and also with the Center for Intelligent Networking and Communications (CINC), University of Electronic Science and Technology of China (UESTC), Chengdu, China (email: linzhang1913@gmail.com).}
\thanks{ Z. Wang is with Ericsson Research, Stockholm, Sweden (email: zhao.wang@ericsson.com).}

\thanks{M. Xiao is with the School of Electrical Engineering, Royal
Institute of Technology (KTH), Stockholm, Sweden (email: mingx@kth.se).}

\thanks{G. Wu and S. Li are with the Key Laboratory on Communications, University of Electronic Science and Technology of China (UESTC), Chengdu, China (emails: wugang99@uestc.edu.cn and lsq@uestc.edu.cn).}

\thanks{Y.-C. Liang is with the Center for Intelligent Networking and Communications (CINC), University of Electronic Science and Technology of China (UESTC), Chengdu, China (email: liangyc@ieee.org).}

\thanks{This work was supported in part by National Natural Science Foundation of China under Grants 61571100, 61631005, 61628103, and 61801101.}

%\thanks{Lin Zhang, Gang Wu, Ying-Chang Liang, and Shaoqian Li are with the National Key Lab of
%Science and Technology on Communications, University of Electronic
%Science and Technology of China, Chengdu, China, emails:
%linzhang1913@gmail.com, wugang99@uestc.edu.cn, liangyc@ieee.org, lsq@uestc.edu.cn; Zhao Wang is with Ericsson Research, Stockholm, Sweden, email: zhao.wang@ericsson.com; Ming Xiao is with the School of Electrical Engineering, Royal
%Institute of Technology, KTH, email: mingx@kth.se.}

\thanks{Part of this paper has been presented in \cite{Lin_orderoptimal_2016} at IEEE CIC/ICCC 2016.}
}\maketitle

 %\\$\star$ NCL, University
%of }
%%%\thanks{Lin Zhang, Gang Wu and Shaoqian Li are with the National Key Lab of Science and Technology on
%Communications, University of Electronic Science and Technology of China, Chengdu, China, emails:
%linzhang1913@gmail.com, \{wugang99, lsq\}@uestc.edu.cn; Ming Xiao is with the School of Electrical
%Engineering, Royal Institute of Technology, KTH, email: mxiao@ieee.org; Ying-Chang Liang is with the Institute
%for Infocomm Research, A*STAR, Singapore, and University of Electronic Science and Technology of China,
%Chengdu, China, email: liangyc@ieee.org.}
%}\maketitle

%\thispagestyle{empty}

\begin{abstract}
We study the decentralized caching scheme in a two-layer network, which includes a sever, multiple helpers, and multiple users. Basically, the proposed caching scheme consists of two phases, i.e, placement phase and delivery phase. In the placement phase, each helper/user randomly and independently selects contents from the server and stores them into its memory. In the delivery phase, the users request contents from the server, and the server satisfies each user through a helper. Different from the existing caching scheme, the proposed caching scheme takes into account the pre-stored contents at both helpers and users in the placement phase to design the delivery phase. Meanwhile, the proposed caching scheme exploits index coding in the delivery phase and leverages multicast opportunities, even when different users request distinct contents. Besides, we analytically characterize the performance limit of the proposed caching scheme, and show that the achievable rate region of the proposed caching scheme lies within constant margins to the information-theoretic optimum. In particular, the multiplicative and additive factors are carefully sharpened to be $\frac{1}{48}$ and $4$ respectively, both of which are better than the state of arts. Finally, simulation results demonstrate the advantage of the proposed caching scheme compared with the state of arts.
\end{abstract}


%\pagestyle{empty}
%\newpage
\section{Introduction}

\emph{Video-on-demand} (VoD) and Internet of Things (IoT) are envisioned to generate massive Internet traffic in the next few years and inevitably causes network congestions \cite{Lin_WCM_2017, IoT_1, IoT_2, IoT_3, Lin_noma}. To deal with this issue, the content caching is proposed by utilizing the storage capacities of network nodes \cite{Dowdy1982,
Korupolu2001, Kanga2002, Baev2008, Borst2010}. In particular, if some contents are pre-stored in a local storage close to a user, these pre-stored contents can be directly accessed by the user. This avoids unnecessary content deliveries from servers and thus releases network congestions. Apparently, this mechanism is able to offer a local caching gain, which is particularly relevant when the local storage is large.

Recently, \emph{Maddah-Ali \&
Niesen} (MAU) introduced index coding into a singer-layer content caching network \cite{MadAliUrs2014}, \cite{MadAliUrs2015}, which includes a server and multiple users. In particular, MAU proposed a centralized caching scheme in \cite{MadAliUrs2014} and a decentralized caching scheme in \cite{MadAliUrs2015}, respectively (We name them as MAU-Centralized scheme and MAU-Decentralized scheme respectively hereafter.). The basic idea of the centralized and decentralized caching schemes is as follows: By viewing the pre-stored contents as the side-information, a coded delivery can be designed to create a \emph{single-layer multicast opportunity} (SMO), even when users request distinct contents. With this idea, both caching schemes are able to reduce the traffic load of the network and reveal coding gains. It is worth noting that both coding gains are shown to be proportional to the aggregation of all the local storage capacities. Thus, the content caching with index coding is able to leverage both local caching gain and global coding gain.

Due to the centralized processing, the MAU-Centralized scheme provides a larger coding gain than the MAU-Decentralized scheme. Nevertheless, the MAU-Centralized scheme is sensitive to the instantaneous network profile and needs redesigns once the instantaneous network profile changes, while the MAU-Decentralized scheme is independent from the instantaneous network profile and thus is more robust than the MAU-Centralized scheme. It is remarkable that the MAU-Decentralized scheme is proved to be with the same order-optimality as the MAU-Centralized scheme, i.e., the same multiplicative and additive factors.

In practical situations, the network topology is usually
tree-like and users have to obtain contents through some intermediate nodes, namely, helpers, from a server. Then, \cite{NKaram2016} considered a two-layer network and intended to minimize the traffic loads of both layers, i.e., the first layer is from the server to the helpers and the second layer is from each helper to its attached users. In particular, \cite{NKaram2016} proposed a generalized caching scheme by exploiting both the SMO of each layer and the \emph{cross-layer multicast opportunity} (CMO) between the server and users. The important observation from \cite{NKaram2016} is that almost no tension exists between the traffic loads of two layers, such that the loads of both layers can be reduced simultaneously. Yet, the problem still remains on whether the traffic loads of both layers can be further reduced, which is both practically and theoretically important. In the rest of the paper, we describe the traffic load with a normalized transmission rate, which will be formally defined later, and we use ``traffic load'' and ``transmission rate'' interchangeably. In particular, a large/small transmission rate corresponds to a large/small traffic load.

\subsection{Main contributions}
In this paper, we study the decentralized caching in a two-layer caching network similar to \cite{NKaram2016}. It is worth noting that, a two-layer caching network can be divided into multiple single-layer caching sub-networks. In particular, the server and the helpers form a single-layer caching sub-network, in which each helper requests multiple contents from the server. Each helper and its attached users form a single-layer caching sub-network, in which each user requests one content from the associated helper. Besides \cite{MadAliUrs2014} and \cite{MadAliUrs2015}, advanced caching schemes in a single-layer caching network have been extensively studied in recent years. These advanced caching schemes can be directly applied into each single-layer caching sub-network, if we treat these single-layer caching sub-networks separately. However, the direct application of these advanced caching schemes may be suboptimal in terms of transmission rates in the network. Instead, we focus on the joint caching design of the two layers. To highlight the contributions, we list the main contents of this paper.

%To achieve this, we first design an S$\&$C caching scheme to leverage both the SMO and CCG. Based on the S$\&$C caching scheme, we develop a hybrid caching scheme to simultaneously exploit the SMO, CCG, and CMO. After that, we study the performance limit of the hybrid caching scheme.

%Basically, the principle of the proposed caching schemes originates from the index coding caching, i.e., MAU-Decentralized scheme, in \cite{MadAliUrs2015}, which is also adopted in \cite{NKaram2016}. Since the index coding caching in \cite{MadAliUrs2015} is proved to be suboptimal in the single-layer caching network, our proposed hybrid caching scheme is also suboptimal in the two-layer caching network. To highlight the contributions, we list the main contents of this paper.

\begin{itemize}
\item We notice that the pre-stored contents at user's storage do not need to be recovered at the associated helper. This reduces the transmission rate of the first layer and creates a \emph{cross-layer caching gain} (CCG). Then, we develop an S$\&$C caching scheme to exploit both the SMO and CCG. Meanwhile, we analytically derive the corresponding transmission rate of each layer with a closed-form expression.

%    By comparing the closed-form expressions with those in \cite{NKaram2016}, we show that, the transmission rate in the first layer can be reduced without increasing the achievable rate in the second layer.

\item We propose a hybrid algorithm to exploit the SMO, CMO, and CCG simultaneously. Specifically, we segment the whole network into two parallel sub-networks and combine the S$\&$C caching scheme and the caching scheme B in
\cite{NKaram2016} in a memory-sharing manner. Meanwhile, we introduce $\alpha, \beta \in [0,1]$ as the memory sharing parameters in the network segmentation, and analytically derive the transmission rate of each layer with a closed-form expression as a function of $\alpha$ and $\beta$. Notably, different tuples of $(\alpha, \beta)$ correspond to different caching designs. By comparing the closed-form expressions with those in \cite{NKaram2016}, we show that the transmission rate of the first layer can be reduced without increasing the transmission rate of the second layer for any $(\alpha, \beta)$. It is worth noting that, the reduction of the transmission rate in the first layer is of significant importance, since the maximum traffic load at a server in the first layer is usually the bottleneck of the network capacity and the load reduction at the server enables the server to support more helpers/users and thus boosts the network capacity.

\item We analytically optimize $(\alpha, \beta)$ and obtain the performance limit of the hybrid caching scheme as follows,
\begin{equation} \nonumber
\mathcal{R}^{H}(M_1,M_2) \subset \mathcal{R}(M_1,M_2) \subset
c_1\mathcal{R}^{H}(M_1,M_2)-c_2, \label{Limits_hybrid}
\end{equation}
where $M_1$ and $M_2$ are the normalized storage sizes at helpers and users respectively,
$\mathcal{R}^{H}$ is the achievable rate region with the proposed
hybrid caching scheme and $\mathcal{R}$ is the information-theoretic (optimal) rate region. In particular, the multiplicative and additive factors are carefully quantified to be $c_1=\frac{1}{48}$ and
$c_2=4$ respectively, both of which are better than those in \cite{NKaram2016}.

\end{itemize}

Compared with \cite{NKaram2016}, the contribution of the paper is four-fold. Firstly, we propose a hybrid caching scheme to reduce the traffic load in a two-layer caching network. Secondly, we provide optimized tuples of $(\alpha, \beta)$ for different settings of the network. Thirdly, we elaborate a better quantification of the gaps between the achievable rates and the information-theoretic minimum rates. Finally, by adopting the overall traffic load in the network as the metric in simulations, we demonstrate that the proposed caching scheme is able to effectively reduce the traffic loads in different network settings compared with the caching scheme in \cite{NKaram2016}. Although the motivation of the paper originates from a trivial observation of the scheme in \cite{NKaram2016}, the proposed caching scheme and the associated information-theoretic analysis are not straightforward.

%Note that, a two-layer caching network can be divided into multiple single-layer caching subnetworks. In particular, the server and helpers form a single-layer caching subnetwork, where each helper requests multiple files from the server. Each helper and the attached users form a single-layer caching subnetwork, where each user requests a file from the helper. After \cite{MadAliUrs2014} and
%\cite{MadAliUrs2015}, advanced caching schemes in the single-layer caching network have been extensively studied in the past few years. If we treat these single-layer caching subnetworks separately, these advanced caching schemes can be applied into each single-layer caching subnetwork to reduce the transmission rates of the server and each helper. In this paper, similar to \cite{NKaram2016}, we develop caching algorithms for a two-layer network based on the decentralized caching scheme in \cite{MadAliUrs2015}. Basically, our main idea is to exploit the correlations of the pre-stored contents at the helpers and the attached users to reduce the transmission rates of the server and each helper, and study the impact of the content correlations on the optimality of the decentralized caching scheme. In fact, our idea is also applicable if we adopt the existing advanced caching schemes for each single-layer subnetwork in the considered two-layer caching network. The extension to the existing advanced caching schemes is beyond the scope of this paper and will not be discussed.

\subsection{Related works}
After \cite{MadAliUrs2014} and
\cite{MadAliUrs2015}, the coded caching problems are widely studied from various aspects. \cite{HooshangAditya2017} studied the centralized coded caching and intended to improve the performance of the caching scheme in \cite{MadAliUrs2014}. In particular, authors in \cite{HooshangAditya2017} treated the design of caching scheme as a combinatorial problem of optimally labeling the leaves of a directed tree. By developing a novel labeling algorithm, the lower bound of the transmission rate in \cite{MadAliUrs2014} is significantly improved. \cite{Mohammad2017} proposed a novel coded caching scheme subject to a storage size constraint, and showed the improvement of the required transmission rate compared with the existing approaches when the number of users satisfied some conditions. The coded caching with different storage sizes is studied in \cite{Mohammad2016,AbdelrahmanWCNC2017}. In particular, \cite{Mohammad2016} proposed a new decentralized coded caching scheme and showed the reduction of the required
transmission rate compared with the existing results when the number of users is larger than the number of files. \cite{AbdelrahmanWCNC2017} proposed an optimization framework for cache placement and delivery schemes, which explicitly accounted for different storage sizes and characterized the optimal caching
scheme when the overall users storage size is no larger than the storage size of the file server.

\cite{UrsMadAli2017} and \cite{Jimingyue2015} considered the randomness of
user demands and intended to reduce the average transmission rate. In particular, \cite{UrsMadAli2017} partitioned files with similar request probabilities into a group, and applied the caching scheme in \cite{MadAliUrs2015} to
each group. However, the scheme in \cite{UrsMadAli2017} cannot guarantee rate order-optimality in all regimes of the system parameters. Then, \cite{Jimingyue2015} optimized the caching scheme in \cite{UrsMadAli2017} and proposed a new caching approach with the theoretically analysis of its order-optimality. \cite{Jinbei2015} considered different file sizes and studied the performance limits of coded caching. In particular, \cite{Jinbei2015} derived a new lower bound and an achievable upper bound for the worst case transmission rate.

Besides, \cite{Ji2014, Mingyue2015,Avik2017} considered multiple demands from each user and \cite{OnlineCaching2013,Qifa2017} investigated online caching schemes. Other researches include caching problems in heterogeneous networks \cite{JadHachem2017,Xuejian2017}, \emph{device-to-device} (D2D) assisted caching \cite{Jimingyue2014,Lin2016}, caching in the finite length regime \cite{Shanmugam2016, Shanmugam2017, Tang2017}, security in caching networks \cite{Avik2015, Vaishakh2016, Frederic2016}, and so on.

%in \cite{Konstantinos2016}, the complexity of the optimal coded caching in hierarchical networks is studied.
%
%Heterogenous content popularity
%is investigated in \cite{Hachem2015}.
%In addition, the \emph{device-to-device} (D2D) assisted content
%caching is considered in \cite{Jimingyue2014}. It is observed that
%single-layer networks are studied among these literature, in which users request files directly from the server.

\section{System Model}

%\begin{figure}
%  \centering
%  % Requires \usepackage{graphicx}
%  \includegraphics[width=140mm]{SysModel}\\
%  \caption{}\label{}
%\end{figure}

\begin{figure}[!t]
\centering
\includegraphics[scale=0.4]{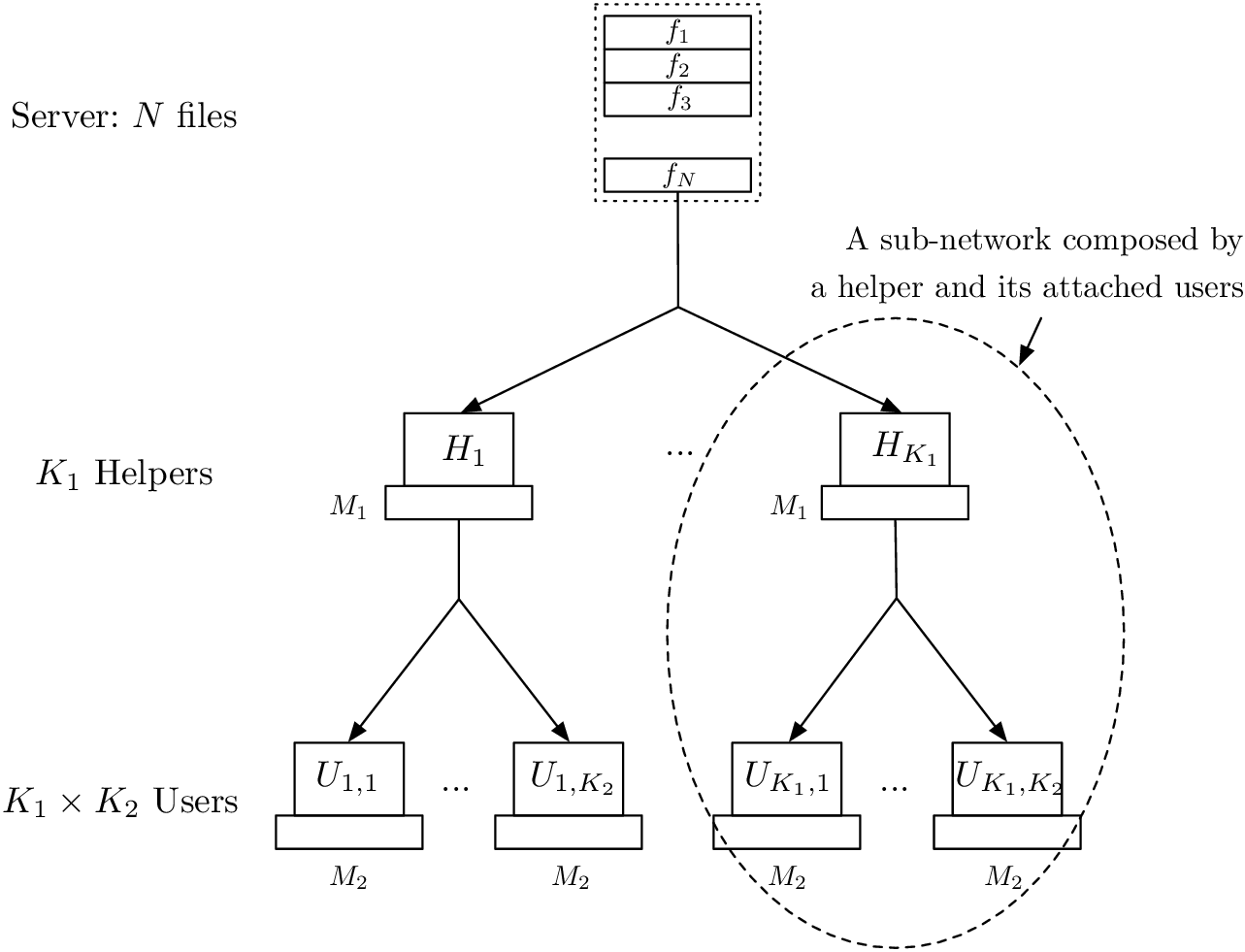}
%\centerline{\epsfig{file=./figure/SysModel.eps,width=140mm}}
\caption{A two-layer network with one server hosting $N$ files, $K_1$ helpers with normalized memory size $M_1$, i.e., $M_1F$ bits, and $K_1K_2$ users with normalized memory size $M_2$, i.e., $M_2F$ bits. Each helper is dedicated for helping $K_2$ users. The straight line between two nodes represents the wireless channel. Similar to [10], we assume that there is no interference among different sub-networks.}
\label{fig:system_model}
\end{figure}

We consider a two-layer caching network as shown in Fig. \ref{fig:system_model}, which includes a server, $K_1$ helpers, and $K_1K_2$ users. In this network, the server is connected to $K_1$ ($K_1\geq 2$) helpers $H_{i}$ ($i \in [1:K_{1}]$), and each helper is exclusively connected to $K_2$ ($K_2\geq 2$)\footnote{For $K_1= 1$ or $K_2=1$, the two-layer network can be
reduced to a single-layer network. Thus, we study the two-layer
network with $K_1\geq 2$ and $K_2\geq 2$ in this paper.} users $U_{i,j}$ ($j \in [1:K_{2}]$).
%Then, we denote the set of helpers as $\mathcal{H}=\{H_i: i \in \{1, 2, \cdots, K_1\}\}$ and the set of users as $\mathcal{U}=\{U_{i, j}| i \in \{1, 2, \cdots, K_1\}, j \in \{1, 2, \cdots, K_2\}\}$.
In particular, the server hosts $N$ files $\mathcal{S}=\{f_n:n\in [1: N]\}$, each of which has $F$ bits. The storage capacities of each helper and each user are $M_1F$ bits and $M_2F$ bits ($M_1, M_2\leq N$)\footnote{For $M_1>N$, it is straightforward to reduce the two-layer system to a single-layer system. For $M_2>N$, each user is able to cache all the files to the local storage.}, respectively. For convenience, we denote $M_1=\frac{M_1F}{F}$ and $M_2=\frac{M_2F}{F}$ as the normalized storage capacities of each helper and each user, respectively. The content caching in the network has two orthogonal phases, i.e., placement phase and delivery phase. In the placement phase, $M_1F$ bits are pre-stored in the storage of each helper and $M_2F$ bits are pre-stored in the storage of each user. In the delivery phase, each user requests a specific file. If some contents of the requested file are in the local storage, the user directly accesses these contents. Otherwise, the user requests these contents from the associated helper and/or server.

%$\mathbf{P}^{H_i} : \mathbb{F}_2^{NF} \rightarrow \mathbb{F}_2^{M_1F}$ ($i\in[1:K_{1}]$), and meanwhile feeds the caches of $K_1K_2$ users by using $K_1K_2$ placement functions $\mathbf{P}^{U_{i,j}}: \mathbb{F}_2^{NF} \rightarrow
%\mathbb{F}_2^{M_2F}$ ($i\in[1:K_{1}], j\in[1:K_{2}]$).

%, the user requests the contents   the server places contents into the memories of $K_1$ helpers by using $K_1$ placement functions  In the delivery phase, each user $U_{i,j}$ requests for a file $f_{d_{i, j}}$ ($d_{i, j} \in [1:N]$) from the server via the corresponding helper $H_i$. After receiving the requested set $\mathcal{D}=\{f_{d_{i, j}}: i \in [1: K_1], j \in [1: K_2]\}$ from $K_1K_2$ users, the server transmits a message $Y_S(\mathcal{D}, \mathbf{P}^{H_i}, \mathbf{P}^{U_{i,j}}: i\in[1:K_{1}], j\in[1:K_{2}])$ (referred as $Y_S$ hereafter) to
%$K_1$ helpers. Sequentially, combining the received message $Y_S$ from the server, the placement functions $\mathbf{P}^{H_i}$ and $\mathbf{P}^{U_{i,j}}$, and the requested set $\mathcal{D}_i=\{f_{d_{i, j}}: j\in [1:K_2]\}$, each helper $H_i$ transmits a message $Y_{H_i}(Y_S, \mathcal{D}_i, \mathbf{P}^{H_i}, \mathbf{P}^{U_{i,j}}: j \in [1: K_2])$ (referred as $Y_{H_i}$ hereafter) to the attached $K_2$ users. Then, each user $U_{i,j}$ recovers the requested file $f_{d_{i,j}}$ from the received message $Y_{H_i}$ and the placement function $\mathbf{P}^{U_{i,j}}$.

It is worth noting that the transmission rates in the network (including the transmission rate from the server to the helpers and the transmission rate from each helper to its attached users) are determined by the requested files of users and the pre-stored (cached) contents in each helper/user. Then, different helpers may have different transmission rates depending on the requested files of users and the pre-stored (cached) contents in each helper/user. Similar to [10], we are interested in the worst case in which different users request distinct files, and focus on the decentralized caching scheme in which each helper/user caches contents randomly and independently\footnote{Note that, although centralized caching may further reduce the traffic load in the two-layer network, it is not straightforward to apply centralized caching into the two-layer network. The reasons are as follows: On one hand, in the single-layer caching network, the centralized caching in [8] is different from the decentralized caching in [9] in terms of content placement schemes, content delivery schemes, and information-theoretic quantifications. On the other hand, the two-layer network is more complicated than the single-layer network, since joint optimizations (including content placement schemes, content delivery schemes, and information-theoretic quantifications) between the two layers need to be conducted. The centralized caching is beyond the scope of the paper.}. Thus, different helpers have an identical (worst-case) rate. In particular, by denoting $L_1$ as the number of the delivered bits from the server to the helpers and denoting $L_2$ as the number of the delivered bits from each helper to its attached users, we define $R_1=\frac{L_1}{F}$ and $R_2=\frac{L_2}{F}$ as the normalized transmission rate from the server and each helper, respectively.

\section{S$\&$C caching scheme: exploiting both SMO and CCG}

In this section, we propose an S$\&$C caching scheme to exploit the SMO from the server to the helpers, the SMO from each helper to its attached users, and the CCG between the helpers and the users. In what follows, we first present the principle and results of this scheme, and then provide the placement algorithm and delivery algorithm, respectively.

\subsection{Principle and results of the S$\&$C caching scheme }

In the considered two-layer caching network, if we directly apply
the MAU-Decentralized scheme in each layer, each requested file will
be resolved at both the user and the associated helper. In fact, if we jointly take the pre-stored contents at the helper and the user into account, the pre-stored contents at the user do not need to be recovered at the associated helper. Therefore, it is possible to reduce the transmission rate with the CCG between helpers and users.

%It is worth noting that, each helper requests $K_2$ files from the server within the considered network. The scheme in \cite{NKaram2016} treats $K_2$ requests of each helper separately and satisfies $K_2$ requests of each helper based on the index coding. Alternatively, the schemes in \cite{Ji2014, Mingyue2015} jointly process $K_2$ requests of each helper and satisfy $K_2$ requests of each helper based on the index coding and the coloring principle in graph theory. On one hand, the achievable rate of the server with the schemes in \cite{Ji2014, Mingyue2015} are smaller than that with the scheme in \cite{NKaram2016} due to a joint processing gain. On the other hand, the introduction of the coloring in \cite{Ji2014, Mingyue2015} may inevitably increase the design complexity. In this paper, although we develop the S$\&$C caching scheme by utilizing the content correlations and treating $K_2$ requests of each helper separately, the key idea of utilizing the content correlations is also applicable by jointly processing $K_2$ requests of each helper, which is beyond the scope of this paper and will not be discussed.

\begin{lemma}
For $M_1\leq N$ and $M_2\leq N$, with the S$\&$C caching scheme, the transmission rate from the server to the helpers is
\begin{align}
R_1^{S\&C}=K_2\left(1-\frac{M_2}{N}\right)\gamma\left(\frac{M_1}{N},K_1\right),
\label{CSC_Rate1}
\end{align}
and the transmission rate from each helper to its attached users is
\begin{equation}
%\begin{split}
R_2^{S\&C}=\gamma\left(\frac{M_2}{N},K_2\right),
\label{CSC_Rate2}
%\end{split}
\end{equation}
where
\begin{align}
\gamma\left(\frac{M}{N},K\right)\triangleq K\left(1-\frac{M}{N}\right)\frac{N}{KM}\left(1-\left(1-\frac{M}{N}\right)^{K}\right)\geq 0.
\label{gamma_I}
\end{align}
\end{lemma}

\begin{remark}[Performance comparison with \cite{NKaram2016}]
In the decentralized caching scheme A of \cite{NKaram2016}, the MAU-Decentralized scheme is directly applied in each individual layer. Accordingly, the transmission rate from the server to the helpers is $R^{A}_1=K_2\gamma\left(\frac{M_1}{N},K_1\right)$ and the transmission rate from each helper to its attached users is $R^{A}_2=\gamma\left(\frac{M_2}{N},K_2\right)$. By comparing $R^{S\&C}_1$ with $R^{A}_1$, we have $R^{S\&C}_1=R^{A}_1\left(1-\frac{M_2}{N}\right)$, which is strictly smaller than $R^{A}_1$ due to $1-\frac{M_2}{N}<1$. Besides, we observe that $R^{S\&C}_2$ is equal to $R^{A}_2$. Therefore, with the S$\&$C caching scheme, the transmission rate in the first layer can be reduced without increasing the transmission rate in the second layer compared with the decentralized caching scheme A in \cite{NKaram2016}.
\end{remark}

\subsection{Placement and Delivery of the S$\&$C scheme}
In this part, we present the detailed placement and delivery methods
of the S$\&$C scheme.

%The pseudo-code of the scheme is provided in \textbf{Algorithm \ref{CSC_scheme}}.

\subsubsection{Placement at helpers and users}

For the placement at helpers, each helper randomly and independently selects $\frac{M_1F}{N}$ bits from each file $n \in [1:N]$ and stores them into its storage. Similarly, for the placement at users, each user randomly and independently selects $\frac{M_2F}{N}$ bits from each file $n \in [1:N]$ and stores them into its storage.

\subsubsection{Delivery from the server to helpers}

To begin with, let $d_{i,j}$ denote the requested file index of $U_{i,j}$, let $V_{d_{i,j},S}$ denote the contents of file $f_{d_{i,j}}$ that are exclusively cached at the nodes in set $S$. From the result in \cite{MadAliUrs2015}, the server needs to transmit $\{\oplus_{i\in S^H_1}V_{d_{i,j},S^H_1\setminus{i}}: S^H_1\subset[1:K_1], |S^H_1|=s_1, s_1\in[1:K_1], j\in [1,K_2]\}$, where $S^H_1$ denotes the set of helper indexes, such that each helper can recover the requested files of its attached users. Since the contents are randomly and independently placed at helpers and users, each bit in $V_{d_{i,j},S^H_1\setminus{i}}$ is cached in user $U_{i,j}$ with probability $\frac{M_2}{N}$. As a result, we can divide each $V_{d_{i,j},S^H_1\setminus{i}}$ in $\{\oplus_{i\in S_1}V_{d_{i,j},S^H_1\setminus{i}}: S^H_1\subset[1:K_1], |S^H_1|=s_1, s_1\in[1:K_1], j\in [1,K_2]\}$ into two parts, i.e.,
\begin{equation}
V_{d_{i,j},S^H_1\setminus{i}}=\{V_{d_{i,j},S^H_1\setminus{i}}^{U_{i,j}}, V_{d_{i,j},S^H_1\setminus{i}}^{\bar U_{i,j}}\},
\end{equation}
where $V_{d_{i,j},S^H_1\setminus{i}}^{U_{i,j}}$ denotes the part that is cached in user $U_{i,j}$ and $V_{d_{i,j},S^H_1\setminus{i}}^{\bar U_{i,j}}$ is the part that is not cached in user $U_{i,j}$. Then, we have
\begin{equation}
|V_{d_{i,j},S^H_1\setminus{i}}^{U_{i,j}}| \approx \frac{M_2}{N}|V_{d_{i,j},S^H_1\setminus{i}}|
\end{equation}
and
\begin{equation}
|V_{d_{i,j},S^H_1\setminus{i}}^{\bar U_{i,j}}| \approx \left(1-\frac{M_2}{N}\right)|V_{d_{i,j},S^H_1\setminus{i}}|.
\end{equation}
In particular, when the file size $F$ is large enough, the approximations in (5) and (6) can be replaced by equalities according to the law of large numbers.

Since $V_{d_{i,j},S^H_1\setminus{i}}^{U_{i,j}}$ can be locally accessed by user $U_{i,j}$, helper $H_i$ does not need to recover $V_{d_{i,j},S^H_1\setminus{i}}^{U_{i,j}}$. In other words, the server only needs to transmit $\{\oplus_{i\in S^H_1}V_{d_{i,j},S^H_1\setminus{i}}^{\bar U_{i,j}}: S^H_1\subset[1:K_1], |S^H_1|=s_1, s_1\in[1:K_1], j\in [1,K_2]\}$, such that helper $H_i$ obtains all the requested subfiles in $f_{d_{i,j}}^{\bar H_i, \bar U_{i,j}}=\{V_{d_{i,j},S^H_1\setminus{i}}^{\bar U_{i,j}}: S^H_1\subset[1:K_1], i\in S_1, |S^H_1|=s_1, s_1\in[1:K_1], j\in [1,K_2]\}$ that are pre-stored in neither of helper $H_i$ and users. Recall that, the transmission rate from the server with the scheme in \cite{MadAliUrs2015} is $R^{A}_1=K_2\gamma(M_1,K_1)$. Thus, the transmission rate from the server with the proposed approach is $K_2\left(1-\frac{M_2}{N}\right)\gamma\left(\frac{M_1}{N},K_1\right)=\left(1-\frac{M_2}{N}\right)R^{A}_1$ as shown in (\ref{CSC_Rate1}).

\subsubsection{Delivery from helper $H_i$ to the attached users}
From the result in \cite{MadAliUrs2015}, helper $H_i$ needs to transmit $\{\oplus_{j\in S^{U}_2}V_{d_{i,j},S^{U}_2\setminus{j}}: S^{U}_2\subset[1:K_2], |S^{U}_2|=s_2, s_2\in[1:K_2]\}$, where $S^{U}_2$ denotes the set of user indexes attached to helper $H_i$, such that each attached user can recover the requested files. Here, we also enable helper $H_i$ to transmit $\{\oplus_{j\in S^{U}_2}V_{d_{i,j},S^{U}_2\setminus{j}}: S^{U}_2\subset[1:K_2], |S^{U}_2|=s_2, s_2\in[1:K_2]\}$ and satisfy its attached users. In what follows, we shall prove that all the contents in $\{\oplus_{j\in S^{U}_2}V_{d_{i,j},S^{U}_2\setminus{j}}: S^{U}_2\subset[1:K_2], |S^{U}_2|=s_2, s_2\in[1:K_2]\}$ have been pre-stored or recovered by helper $H_i$.

According to whether the contents in $V_{d_{i,j},S^{U}_2\setminus{j}}$ are pre-stored in helper $H_i$, $V_{d_{i,j},S^{U}_2\setminus{j}}$ can be divided into two parts, i.e.,
\begin{equation}
V_{d_{i,j},S^{U}_2\setminus{j}}=\{V_{d_{i,j},S^{U}_2\setminus{j}}^{H_i}, V_{d_{i,j},S^{U}_2\setminus{j}}^{\bar H_i}\},
\end{equation}
where $V_{d_{i,j},S^{U}_2\setminus{j}}^{H_i}$ denotes the part that is pre-stored in helper $H_i$ and $V_{d_{i,j},S^{U}_2\setminus{j}}^{\bar H_i}$ is the part that is not pre-stored in helper $H_i$. It is clear that $V_{d_{i,j},S^{U}_2\setminus{j}}^{\bar H_i}$ is pre-stored in neither of helper $H_i$ nor user $U_{i,j}$. Then, $V_{d_{i,j},S^{U}_2\setminus{j}}^{\bar H_i}$ must be in  $f_{d_{i,j}}^{\bar H_i, \bar U_{i,j}}$, which has been obtained by helper $H_i$. Thus, if helper $H_i$ transmits $\{\oplus_{j\in S^{U}_2}V_{d_{i,j},S^{U}_2\setminus{j}}: S^{U}_2\subset[1:K_2], |S^{U}_2|=s_2, s_2\in[1:K_2]\}$, all the requests of its attached users can be satisfied. Recall that, the transmission rate of each helper with the scheme in \cite{MadAliUrs2015} is $R^{A}_2=\gamma\left(\frac{M_2}{N},K_2\right)$. Thus, the transmission rate of each helper with the proposed approach is $R^{S\&C}_2=R^{A}_2=\gamma\left(\frac{M_2}{N},K_2\right)$.

\section{Hybrid caching in two-layer networks}
In this section, we present a hybrid caching scheme to exploit SMO, CMO, and CCG simultaneously. For convenience, we first introduce the caching scheme B of \cite{NKaram2016}, which exploits CMO. Then, we combine the S$\&$C caching scheme with the caching scheme B of \cite{NKaram2016} in a memory-sharing manner and develop the hybrid caching scheme.

\subsection{Caching scheme B in \cite{NKaram2016}}

\subsubsection{Basic principle and results}
To exploit the CMO, the MAU-Decentralized caching scheme is directly applied between the server and $K_1K_2$ users by ignoring the storages of the helpers. Specifically, each user randomly and independently selects $\frac{M_2F}{N}$ bits of each file and stores them into its storage in the placement phase. Then, the server conducts MAU-Decentralized delivery to satisfy all the users through the helpers. Note that there is no direct link between the server and the users. Thus, the server first transmits contents to the helpers. Then, each helper forwards the contents, which are relevant to the requested files of its attached users. Finally, each user recovers the requested file by combining the forwarded contents of the associated helper and the pre-stored contents in its own storage. With this caching scheme, the transmission rate from the server to the helpers is
\begin{equation}
%\begin{split}
R_1^{B}=\gamma\left(\frac{M_2}{N}, K_1K_2\right)
\label{B_Rate1}
%\end{split}
\end{equation}
and the transmission rate from each helper to its attached users is
\begin{equation}
%\begin{split}
R_2^{B}=\gamma\left(\frac{M_2}{N}, K_2\right).
\label{B_Rate2}
%\end{split}
\end{equation}

\subsubsection{Placement at users}
In the placement phase, each user randomly and independently selects $\frac{M_2F}{N}$
bits of each file $n \in [1:N]$ and caches them into its storage.

\subsubsection{Delivery from server to helpers}
From the results in \cite{MadAliUrs2015}, the server needs to transmit
$\{\oplus_{(i-1)K_2+j \in S_3}V_{d_{i,j}, S_3\setminus{(i-1)K_2+j}}:
S_3\subset [1:K_1K_2], |S_3|=s_3, s_3\in[1:K_1K_2]\}$, where $S_3$ denotes a set of user indexes, such that each user can recover the requested file. Then, the transmission rate from
the server to the helpers can be obtained as (\ref{B_Rate1}).

\subsubsection{Delivery from helper $H_i$ to the attached users}
From the results in \cite{MadAliUrs2015}, if helper $H_i$ forwards the contents
relevant to the requested files of its attached users, the transmission rate from helper $H_i$ to its attached users can be obtained as (\ref{B_Rate2}).

\subsection{Hybrid caching scheme}

%\begin{figure}[!t]
%%\centerline{\epsfig{file=generalized.eps,width=80mm}}
%            \centering
%            \includegraphics[scale=0.7]{generalized}
%            \caption{Hybrid caching scheme. For given $\alpha$ and $\beta$, we split the system into two subsystems. In particular, we use S$\&$C caching scheme to deliver the first parts of the files over the first subsystem and use the CMO scheme to deliver the second parts of the files over the second subsystem.}
%            \label{generalized}
% \end{figure}

%In this section, we introduce another joint caching scheme, i.e., the H-C scheme. Different from the H-C scheme and the H-C scheme, the H-C scheme combines the two schemes in a memory-sharing manner, such that both the memories and multicast opportunities across two layers are elaborated. In what follows, we first give the principle of the H-C scheme. Then, we provide the performance of this scheme.

\subsubsection{Basic principle and results}
Note that, the S$\&$C caching scheme creates both SMO and CCG by utilizing the storage capacity of the helpers, while the decentralized caching scheme B of \cite{NKaram2016} creates CMO by ignoring the storage capacity of each helper. To exploit SMO, CCG, and CMO simultaneously, we divide the whole network into two parrel sub-networks and apply S$\&$C caching scheme and the decentralized caching scheme B of \cite{NKaram2016} in a memory-sharing manner. In particular, we split each file into two parts with sizes $\alpha F$ ($ 0 \leq
\alpha \leq 1$) and $(1-\alpha)F$ bits. Meanwhile, we partition
the storage capacity of each user into two parts of normalized storage capacities $\beta M_2$ ($ 0 \leq
\beta \leq 1$) and $(1-\beta)M_2$. Then, the first sub-network consists
of a server hosting the first fraction $\alpha$ part of each file,
$K_1$ helpers each with a normalized storage capacity $M_1$, and $K_1K_2$ users
each with a normalized storage capacity $\beta M_2$. The second sub-network
consists of the server hosting the second fraction $1-\alpha$ part
of each file, $K_1$ helpers each without any storage capacity, and $K_1K_2$
users each with a normalized storage capacity $(1-\beta)M_2$. In this way, we can apply the S$\&$C caching scheme in the first sub-network which delivers the first fraction $\alpha$ part of each file, and apply the CMO caching
scheme in the second sub-network which delivers the second fraction $1-\alpha$ part of each file. It is worth noting that, with the hybrid caching scheme, the overall transmission rate of each layer is the sum rate of the layer in two sub-networks.

%It should be noted that, to utilize the whole
%memories in helpers and users, the values of $\alpha$ and $\beta$
%need to satisfy $\alpha N \geq M_1$, $M_1\geq \beta M_2$, and
%$(a-\alpha)N\geq (1-\beta)M_2$, i.e., $\frac{M_1}{N} \leq \alpha
%\leq 1$, $0 \leq \alpha \leq \min{1, \frac{M_1}{M_2}}$, and $\alpha
%N-\beta M_2 \leq N-M_2$. For illustration, we provide the region of
%$(\alpha, \beta)$ in Fig. \ref{alpha_beta}.
%
%\begin{figure*}[h]
%\centerline{\epsfig{file=alpha_beta.eps,width=80mm}}
%%            \centering
%%            \includegraphics[scale=0.8]{generalized}
%            \caption{The region of $(\alpha, \beta)$ that enables to utilize the whole memories in helpers and users.}
%            \label{alpha_beta}
% \end{figure*}

\begin{lemma} For $M_1\leq N$ and $M_2\leq N$, the transmission rate from the server to the helpers is
\begin{align}\nonumber
R_1^{H}(\alpha, \beta)=&\alpha K_2 \left[\left(1-\frac{\beta M_2}{\alpha N}\right)\right]^+ \left[\gamma\left(\frac{M_1}{\alpha N},K_1\right)\right]^+ +\\
&(1-\alpha)\left[\gamma\left(\frac{(1-\beta)M_2}{(1-\alpha)N},K_1K_2\right)\right]^+
\label{Hybrid_Rate1}
\end{align}
and the transmission rate from each helper to its attached users is
\begin{align} \nonumber
R_2^{H}(\alpha, \beta)=&\alpha \left[\gamma\left(\frac{\beta M_2}{\alpha N},K_2\right)\right]^++\\
&(1-\alpha)\left[\gamma\left(\frac{(1-\beta)M_2}{(1-\alpha)N},K_2\right)\right]^+,
\label{Hybrid_Rate2}
\end{align}
where $[x]^+\triangleq \max\{x,0\}$.
\end{lemma}
\begin{proof}
$R_1^{H}(\alpha, \beta)$ and $R_2^{H}(\alpha, \beta)$ can be obtained by applying the S$\&$C caching scheme and the caching scheme B of \cite{NKaram2016} in a memory-sharing manner with factor $\alpha$ and $\beta$. The proof is similar to that in \cite{NKaram2016} and will be omitted for space limitation.
\end{proof}

\begin{remark}
\emph{Performance comparison with \cite{NKaram2016}:} It is worth noting that, the generalized caching scheme in \cite{NKaram2016} is developed by combining the caching scheme A and B therein in a memory-sharing manner, and the proposed hybrid caching scheme is obtained by combining the S$\&$C caching scheme and the caching scheme B in \cite{NKaram2016}. Recall that, with the S$\&$C caching scheme, the transmission rate in the first layer can be reduced without increasing the transmission rate in the second layer compared with the caching scheme A in \cite{NKaram2016}. Thus, for any $(\alpha, \beta)$, the first-layer transmission rate of the hybrid caching scheme is smaller than that of the generalized caching scheme in \cite{NKaram2016}. In fact, the reduction of the transmission rate in the first layer is of significant importance, since the maximum traffic load at a server in the first layer is usually the bottleneck of the network capacity and the load reduction at the server enables the server to support more helpers/users and thus boosts the network capacity. Meanwhile, the transmission rates in the second layer with both schemes are identical. This is due to the fact that, both the proposed hybrid caching scheme and generalized caching scheme in \cite{NKaram2016} adopt the single-layer caching scheme in \cite{MadAliUrs2015} for the content transmission from each helper to the attached users.
%For illustration, we compare the achievable transmission rates of
%the hybrid caching scheme and the generalized caching scheme in Fig.
%\ref{Comparison_alpha_beta}, where $N=50$, $M_1=10$, $M_2=20$,
%$K_1=10$, and $K_2=2$. Since the proposed hybrid scheme is able to reduce the transmission rate in the first layer and achieves the same transmission rate in the second layer, we only compare the transmission rates in the first layer with two schemes. In Fig. \ref{Comparison_alpha_beta}-(a), we
%set $\beta=0.5$ and increase $\alpha$ from $0.2$ to $0.9$. In Fig.
%\ref{Comparison_alpha_beta}-(b), we set $\alpha=0.5$ and increase
%$\beta$ from $0.2$ to $0.9$. From this figure, given $\alpha$ ($\beta$), the transmission rates of the server can be significantly reduced by the proposed hybrid caching scheme, especially when $\beta$ ($\alpha$) is large. The result shows the advantages of the proposed hybrid caching scheme over the generalized caching scheme in \cite{NKaram2016}.
\end{remark}

%\begin{figure}[!t]
%\centering
%\includegraphics[scale=0.5]{Comparison_alpha_beta}
%\caption{Comparison of the achievable transmission rates of the
%hybrid caching scheme and the generalized caching scheme, where
%$N=50$, $M_1=10$, $M_2=20$, $K_1=10$, and $K_2=2$. In sub-figure
%(a), $\beta=0.5$. In sub-figure (b), $\alpha=0.5$.}
%\label{Comparison_alpha_beta}
%\end{figure}

%Lemma 3 indicates that the transmission rates in
%(\ref{Hybrid_Rate1}) and (\ref{Hybrid_Rate2}) can be achieved
%simultaneously for given $(\alpha, \beta)$. The performance limits
%of $R_{1}^{H}$ and $R_{2}^{H}$ are investigated in the following.

\subsubsection{Performance limit}
Clearly, different tuples of $(\alpha, \beta)$ lead to different hybrid caching designs and transmission rates in two layers. In this part, we first optimize $(\alpha, \beta)$ and evaluate the achievable rates by applying the optimized $(\alpha, \beta)$ into (\ref{Hybrid_Rate1}) and (\ref{Hybrid_Rate2}). Then, we quantify the gap between the achievable rate region and the information-theoretic (optimal) rate region, after characterizing the differences between the achievable rates and their information-theoretic lower bounds in two layers, respectively.

To begin with, we formally define the achievable rate region of the hybrid caching scheme and the information-theoretic (theoretically optimal) rate region in the following.

%For given $K_{1}, K_{2}$ and an arbitrary given requested set $\mathcal{D}$, the tuple $(M_{1}, M_{2}, R_{1}, R_{2})$ is achievable, if for a large enough file size $F$, each requested $f_{d_{i,j}} \in \mathcal{D}$ can be recovered with error probability arbitrarily close to $0$. Our major interest is the \emph{feasible rate region}, which can be defined as follows:
%
%\begin{definition}[Feasible rate region]
%\label{Def:AchieveRateRegion}
%For memory sizes $M_{1}$, $M_{2} \geq 0$, and the number of helpers $K_{1}$, the number of attached users $K_{2}$, the feasible rate region $\mathcal{R}(M_1,M_2)$ is defined as the closure of rate pairs $(R_{1}, R_{2})$, such that $(M_{1}, {M}_{2}, R_{1}, R_{2})$ is achievable.
%\end{definition}

\begin{definition} For normalized memory size $M_1,M_2\geq 0$, we define
\begin{equation}
\mathcal{R}^{H}(M_1,M_2) = \{(R_1^{H}(\alpha, \beta),
R_2^{H}(\alpha, \beta)):\alpha, \beta \in [0,1]\}+\mathbb{R}_+^2
\label{R_Hybrid}
\end{equation}
as the achievable rate region of the hybrid caching scheme, where
$\mathbb{R}_+^2$ denotes the positive quadrant and the addition
corresponds to the Minkowski sum between sets.
\end{definition}

\begin{definition} For normalized memory size $M_1,M_2\geq 0$, we define
\begin{equation}
\mathcal{R}(M_1,M_2) = \{R_1^{lb}(M_1, M_2),
R_2^{lb}(M_1, M_2)\}+\mathbb{R}_+^2
\label{R_Hybrid}
\end{equation}
as the information-theoretic (optimal) rate region, where $R_1^{lb}(M_1, M_2)$ and $R_2^{lb}(M_1, M_2)$ are the information-theoretic lower bounds of transmission rates in two layers, respectively.
\end{definition}

\begin{figure}[!t]
\centering
\includegraphics[scale=0.8]{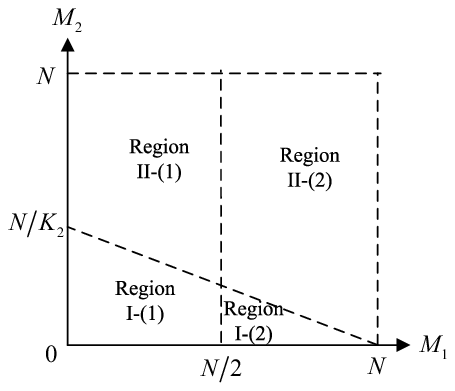}
%\centerline{\epsfig{file=Different_regions.eps,width=80mm}}
\caption{Different regions of ($M_1$,$M_2$). }
\label{Different_regions}
\end{figure}

Due to the complicated expressions of the transmission rates in two layers, it is difficult to obtain the optimal $(\alpha, \beta)$ and minimize the transmission rates in two layers simultaneously. Instead, we adopt a heuristic algorithm to optimize $(\alpha, \beta)$ with the following four steps.

\begin{enumerate}
\item We observe that the aggregated storage size of each sub-network (including the storage size of each helper and the storage size of each attached user) is crucial to the transmission rate/traffic load in the considered network. Then, we divide the feasible region of $(M_1, M_2)$ into two regions: Region I ($M_1+K_2M_2<N$) and Region II ($M_1+K_2M_2\geq N$). In particular, $M_1+K_2M_2<N$ refers to the region in which the aggregated storage size of each sub-network is smaller than the overall size of the files in the server, and $M_1+K_2M_2\geq N$ refers to the region in which the aggregated storage size of each sub-network is no smaller than the overall size of the files in the server.

\item We select multiple tuples of $(\alpha,
\beta)$ with typical values in region I ($M_1+K_2M_2<N$) and evaluate the
corresponding transmission rates via (\ref{Hybrid_Rate1}) and (\ref{Hybrid_Rate2}). The detailed selection of multiple tuples of $(\alpha, \beta)$ is as follows: In Region I, the aggregated storage size of each sub-network is smaller than the overall size of the files in the server. Then, we need to carefully design $(\alpha, \beta)$ and make use of the small caches at helpers/users. In particular, there may exist three kinds of multi-cast opportunities in the considered network. The first one is the multi-cast opportunity from the server to the helpers (namely, MO-1). The second one is the multi-cast opportunity from the server to the users (namely, MO-2). The third one is the multi-cast opportunity from each helper to its attached users (namely, MO-3). Ideally, we need to exploit the three kinds of multi-cast opportunities simultaneously to minimize the traffic load in the network. However, it is challenging to conduct the theoretical analysis when exploiting the three kinds of multi-cast opportunities simultaneously. Instead, we choose typical tuples of $(\alpha, \beta)$ such that two kinds of multi-cast opportunities can be exploited simultaneously. According to [10], $(\alpha, \beta)=(\frac{M_1}{N}, \frac{M_1}{N})$ is able to exploit MO-2 and MO-3 simultaneously, and $(\alpha, \beta)=(\frac{M_1}{M_1+K_2M_2}, 0)$ is able to exploit MO-1 and MO-2 simultaneously. Besides, it is easy to verify that $(\alpha, \beta)=(1, 1)$ can exploit MO-1 and MO-3 simultaneously. Therefore, to make use of the small caches at helpers/users in a flexible manner, we choose three tuples of $(\alpha, \beta)$ for Region I as
      \begin{align}
(\alpha, \beta)=\left\{
\begin{aligned}
&\left(\frac{M_1}{N}, \frac{M_1}{N}\right), \ \text{Tuple I}, \\
& \left(\frac{M_1}{M_1+K_2M_2}, 0\right), \ \text{Tuple II}, \\
&(1, 1), \ \text{Tuple III}.
\end{aligned}
\right.
\end{align}

\item We select multiple tuples of $(\alpha,
\beta)$ with typical values in region II ($M_1+K_2M_2\geq N$) and evaluate the
corresponding transmission rates via (\ref{Hybrid_Rate1}) and (\ref{Hybrid_Rate2}). The detailed selection of multiple tuples of $(\alpha, \beta)$ is as follows: In Region II, the aggregated storage size of each sub-network is no smaller than the overall size of the files in the server. According to [10], for a small value of $\frac{M_1}{N}$, $\left(\alpha, \beta\right)=\left(\frac{M_1}{N},\frac{M_1}{N}\right)$ is able to balance the traffic loads (i.e., $R_1^{H}(\alpha, \beta)$ and $R_2^{H}(\alpha, \beta)$) in the two layers. Then, we choose $\left(\alpha, \beta\right)=\left(\frac{M_1}{N},\frac{M_1}{N}\right)$ as a candidate tuple of $\left(\alpha, \beta\right)$ for a small $\frac{M_1}{N}$. For a relatively large value of $\frac{M_1}{N}$, if we still choose $\left(\alpha, \beta\right)=\left(\frac{M_1}{N},\frac{M_1}{N}\right)$, the traffic load (i.e., $R_1^{H}(\alpha, \beta)$) from the server to helpers may be unacceptably large since $R_1^{H}(\alpha, \beta)$ increases as $\beta$ grows. Thus, for a relatively large value of $\frac{M_1}{N}$, we choose $\left(\alpha, \beta\right)=\left(\frac{M_1}{N},\frac{1}{2}\right)$ to balance the traffic loads in the two layers. To summarize, we choose two tuples of $(\alpha, \beta)$ for Region II as
      \begin{align}
(\alpha, \beta)=\left\{
\begin{aligned}
&\left(\frac{M_1}{N}, \frac{M_1}{N}\right), \ \text{Tuple I}, \\
& \left(\frac{M_1}{N}, \frac{1}{2}\right), \ \text{Tuple II}.
\end{aligned}
\right.
\end{align}

\item We observe that the gap between the achievable rate region $\mathcal{R}^{H}(M_1,M_2)$ and the information-theoretic (optimal) rate region $\mathcal{R}(M_1,M_2)$ is dominated by the difference between the achievable rate and its information-theoretic lower bound in the first layer from the server to the helpers. To reduce the gap between the achievable rate region $\mathcal{R}^{H}(M_1,M_2)$ and the information-theoretic (optimal) rate region $\mathcal{R}(M_1,M_2)$, we calculate the transmission rates of the two layers with the three tuples of $(\alpha, \beta)$ in (14) and choose the tuple that minimizes $R_1^{H}(\alpha, \beta)$ as an optimized design of $(\alpha, \beta)$, i.e., $(\alpha^*, \beta^*)$, for Region I. Similarly, we calculate the corresponding transmission rates of the two layers with the two tuples in (15) of $(\alpha, \beta)$ and choose the tuple that minimizes $R_1^{H}(\alpha, \beta)$ as an optimized design of $(\alpha, \beta)$, i.e., $(\alpha^*, \beta^*)$, for Region II.
\end{enumerate}

To this end, we obtain an optimized $(\alpha^*, \beta^*)$, and the corresponding transmission rates $R_1^{H}(\alpha^*,\beta^*)$ and $R_2^{H}(\alpha^*,\beta^*)$ in each region. With the optimized $(\alpha^*, \beta^*)$, we can characterize the gap between the achievable rate region of the proposed hybrid caching scheme and the information-theoretic rate region in the following theorem.

\begin{theorem}
For $M_1\leq N$ and $M_2\leq N$, we have
\begin{equation}
\mathcal{R}^{H}(M_1,M_2) \subset \mathcal{R}(M_1,M_2) \subset \frac{1}{48}\mathcal{R}^{H}(M_1,M_2)-4.
\label{Limits_hybrid}
\end{equation}
\end{theorem}

\begin{proof}[Sketches of proof]
Since it is straightforward to obtain $\mathcal{R}^{H}(M_1,M_2) \subset \mathcal{R}(M_1,M_2)$, we only prove $\mathcal{R}(M_1,M_2) \subset \frac{1}{48}\mathcal{R}^{H}(M_1,M_2)-4$.

%Since it is straightforward to obtain $\mathcal{R}^{JC-H}(M_1,M_2) \subset
%\mathcal{R}^*(M_1,M_2)$. Then, we focus on the proof of $\mathcal{R}^*(M_1,M_2) \subset
%c_1 \cdot \mathcal{R}^{JC-H}(M_1,M_2)-c_2$.

Firstly, we let $R_1^{lb}(M_1,M_2)$ and $R_2^{lb}(M_2)$ represent the information-theoretic lower bounds of transmission rates in two layers, respectively. Based on the cut-set theorem \cite{Gamal2011}, we have \cite{NKaram2016}:
\begin{equation}
R_1^{lb}(M_1,M_2)\triangleq \underset{\stackrel{s_1\in \{1, 2, \cdots, K_1\}}{s_2\in \{1, 2, \cdots, K_2\}} }{\max} \frac{s_1s_2(N-s_1M_1-s_1s_2M_2)}{N+s_1s_2}
\label{R_1_lb}
\end{equation}
and
\begin{equation}
R_2^{lb}(M_2)\triangleq \underset{t \in \{1, 2, \cdots, K_2\}}{\max} \frac{t(N-tM_2)}{N+t}.
\label{R_2_lb}
\end{equation}

Secondly, we denote the upper bounds of transmission rates $R_1^{H}(\alpha^*, \beta^*)$ and
$R_2^{H}(\alpha^*, \beta^*)$ as $R_1^{ub}(M_1,M_2)$ and $R_2^{ub}(M_2)$, respectively. Then, we calculate the upper bounds in Appendix A and prove that $R_1^{lb}(M_1,M_2) \geq \frac{1}{48}
R_1^{ub}(M_1,M_2)-4$ and $R_2^{lb}(M_2)\geq
\frac{1}{48}R_2^{ub}(M_2)-4$ can be satisfied simultaneously in Appendix B and Appendix C, respectively. More specifically, we divide Region I and Region II into two
sub-regions based on the value of $M_1$, i.e., whether $M_1$ is
larger or smaller than $N/2$. Therefore, we have four sub-regions in
total to investigate, i.e.,
\begin{align}
\left\{
\begin{aligned}
&M_1\!+\!K_2M_2\!<\!N \ \text{and}\ M_1\!<\!N/2, \ \text{Sub-region I-(1)}, \\
& M_1\!+\!K_2M_2\!<\!N\ \text{and} \ M_1\!\geq\! N/2, \ \text{Sub-region I-(2)}, \\
& M_1\!+\!K_2M_2\!\geq\! N \ \text{and} \ M_1 \!<\! N/2, \ \text{Sub-region II-(1)}, \\
& M_1\!+\!K_2M_2\!\geq\! N \ \text{and} \ M_1\!\geq \!N/2, \ \text{Sub-region II-(2)}.
\end{aligned}
\right.
\end{align}
After characterizing the gap between $R_1^{lb}(M_1,M_2)$ and
$R_1^{ub}(M_1,M_2)$, and the gap between $R_2^{lb}(M_2)$ and
$R_2^{ub}(M_2)$ in the four regions, we summarize that
$R_1^{lb}(M_1,M_2) \geq \frac{1}{48} R_1^{ub}(M_1,M_2)-4$ and
$R_2^{lb}(M_2)\geq \frac{1}{48}R_2^{ub}(M_2)-4$ can be
achieved simultaneously for all possible $(M_1,M_2)$.
\end{proof}

\begin{remark}[Performance comparison with \cite{NKaram2016}]
Similar to the definition in (\ref{R_Hybrid}), we define $\mathcal{R}^{G}(M_1,M_2)$ as the achievable rate region of the generalized caching scheme in \cite{NKaram2016}. Accordingly to the results in \cite{NKaram2016}, we have
\begin{equation}
 \mathcal{R}^{G}(M_1,M_2) \subset \mathcal{R}(M_1,M_2) \subset \frac{1}{60}\mathcal{R}^{G}(M_1,M_2)-16.
\label{Limits_G}
\end{equation}
By comparing the multiplicative and additive factors in (\ref{Limits_hybrid}) and (\ref{Limits_G}), the achievable rate region of the proposed hybrid caching scheme is better than that of the generalized caching scheme in \cite{NKaram2016}. The improvement results from three aspects. The first aspect is the utilization of the CCG, which leads to smaller achievable rates. The second aspect is the optimization of $\alpha$ and $\beta$. The third aspect is a better quantification of the information-theoretic gap, including a better division of the entire region in Fig. 2 and an improved mathematical proof. It should be pointed out that the improvement of the multiplicative and additive factors is obtained by exploiting the gains in the three aspects simultaneously, and any absence of the gains in the three aspects may fail to achieve the improvement. For instance, we would not be able to obtain the same improvement with the division of the subregions in [10], even if we exploit the former two gains.
\end{remark}

\begin{remark}
\emph{(Optimized division of the entire region in Fig. 2 compared with \cite{NKaram2016})}: Basically, the division of the entire region in Fig. 2 is optimized based on the following two observations. On one hand, we observe that the multiplicative factor is mainly determined by the gap between the achievable rate region and the information-theoretic (optimal) rate region in Region I. To optimize the multiplicative factor, we further divide Region I into two sub-regions, i.e., Region I-(1) and Region I-(2). On the other hand, we observe that the additive factor is mainly determined by the gap between the achievable rate region and the information-theoretic (optimal) rate region in Region II. To obtain a better additive factor, we optimize the division of Region II compared with [10]. It is worth pointing out that, dividing the entire region in Fig. 2 into more sub-regions may further improve the quantifications of the multiplicative and additive factors. Nevertheless, this requires a more complicated analysis and will be one of our future works.

%\emph{(Joint optimization of $\alpha$ and $\beta$ compared with \cite{NKaram2016})}: In \cite{NKaram2016}, a pair of $(\alpha, \beta)$ is firmly chose in each region to balance the
%transmission rates in two layers and minimize the gap between the
%achievable rate region $\mathcal{R}^G$ of the generalized caching
%scheme and the information-theoretic (optimal) rate region $\mathcal{R}$. In our design, we
%prepare multiple tuples of $(\alpha, \beta)$ and choose the one
%subject to the minimum $R_1^{ub}(M_1,M_2)$. In this way, we
%reduce the gap between $R_1^{ub}(M_1,M_2)$ and $R_1^{lb}(M_1,M_2)$.
\end{remark}

%\begin{remark}
%\emph{(Optimization of information-theoretic proofs compared with \cite{NKaram2016})}: We observe that the gap between the upper bound $R_1^{ub}(M_1,M_2)$ and the lower bound $R_1^{lb}(M_1,M_2)$ is dominated by the characterization of the multiplicative and additive factors in Region I, i.e., $M_1+K_2M_2<N$. By further carefully dividing Region I into two subregions, we quantify the gap between the upper bound $R_1^{ub}(M_1,M_2)$ and the lower bound $R_1^{lb}(M_1,M_2)$ with better multiplicative and additive factors in Region I compared with results in \cite{NKaram2016}. Thus, we have four subregions instead of three subregions in \cite{NKaram2016}. Specifically, both Region I and Region II are divided into two subregions as shown in Fig. \ref{Different_regions}. It is worth pointing out that, dividing the region in Fig. 2 into more sub-regions requires more complicated information-theoretic proofs. Thus, it is not straightforward to further improve the performance by dividing the region in Fig. \ref{Different_regions} into more sub-regions. Further optimization is one of our future work.
%\end{remark}

\section{Simulation results}
In this section, we provide simulation results to show the advantage of the proposed caching scheme over the caching scheme in [10]. In particular, we adopt the overall traffic load in the network, i.e., $R_1+K_1R_2$, as the metric. Next, we first show the traffic load comparison with general values of $\alpha$ and $\beta$, and then give the traffic load comparison with optimized $\alpha$ and $\beta$ in (14) and (15).

\begin{figure}[!t]
\centering
\includegraphics[scale=0.6]{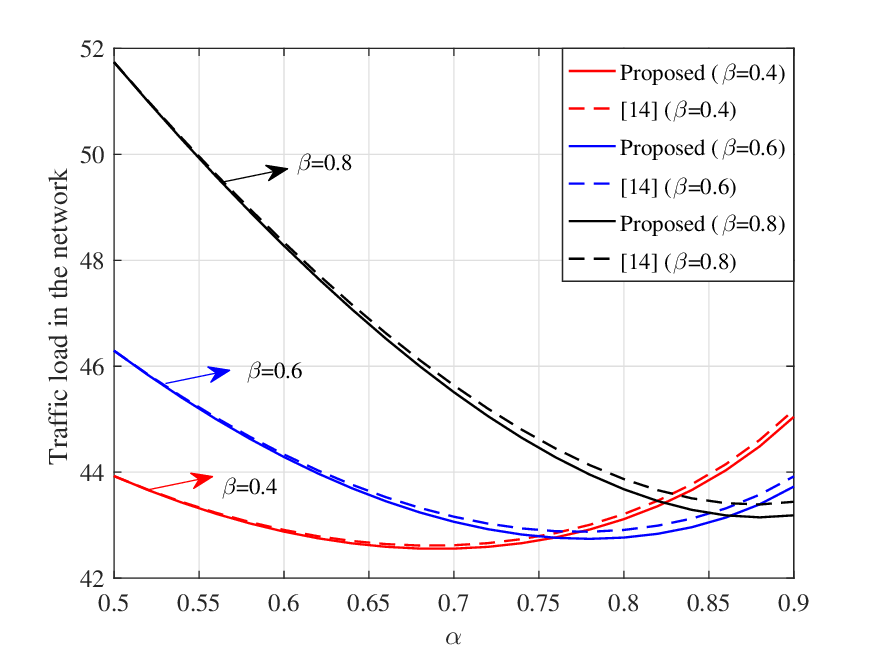}
%\centerline{\epsfig{file=Different_regions.eps,width=80mm}}
\caption{Comparison of the traffic load in the network in Region I with different $\alpha$ and $\beta$.}
\label{Traffic_load_comparison_Region_I}
\end{figure}

\begin{figure}[!t]
\centering
\includegraphics[scale=0.6]{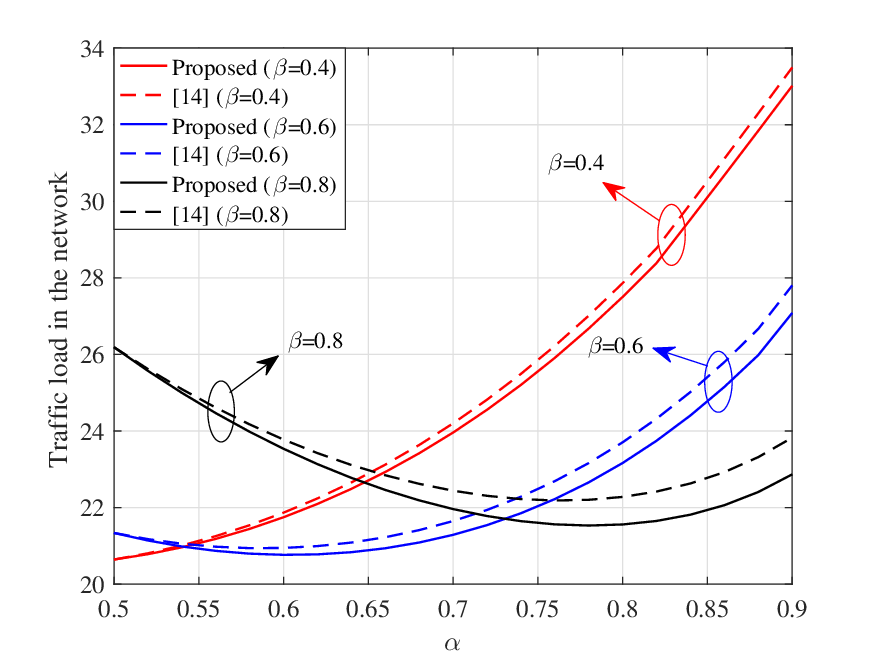}
%\centerline{\epsfig{file=Different_regions.eps,width=80mm}}
\caption{Comparison of the traffic load in the network in Region II with different $\alpha$ and $\beta$.}
\label{Traffic_load_comparison_Region_II}
\end{figure}

Fig. \ref{Traffic_load_comparison_Region_I} and Fig. \ref{Traffic_load_comparison_Region_II} illustrate the traffic load comparison with general values of $\alpha$ and $\beta$ in Region I ($M_1+K_2M_2<N$) and Region II ($M_1+K_2M_2\geq N$), respectively. In particular, we consider the network setting in Region I as follows: $F=1$ M bits, $N=100$, $K_1=10$, $K_2=5$, $M_1=50$, and $M_2=8$. Similarly, we consider the network setting in Region II as follows: $F=1$ M bits, $N=100$, $K_1=10$, $K_2=5$, $M_1=50$, and $M_2=30$. Both figures show that the proposed caching scheme achieves a traffic load similar to the caching scheme in \cite{NKaram2016} for a small $\alpha$, and the proposed caching scheme is able to effectively reduce the traffic load compared with the caching scheme in \cite{NKaram2016} for a relatively large $\alpha$. Besides, the traffic load gap of the two caching schemes becomes large as $\alpha$ increases. The reason is as follows: On one hand, a small $\alpha$ leads to a low CCG of the proposed caching scheme (this can be easily verified from (10)), which has little impact on the traffic load reduction compared with the caching scheme in \cite{NKaram2016}; On the other hand, a large $\alpha$ leads to a high CCG, which contributes to a significant traffic load reduction compared with the caching scheme in \cite{NKaram2016}. Furthermore, by comparing the two figures, we observe that the proposed caching scheme is able to provide larger reductions of traffic load in Region II than in Region I. This is because, the traffic load reduction is proportional to $M_2$, which is small in Region I and is relatively large in Region II.

\begin{figure}[!t]
\centering
\includegraphics[scale=0.6]{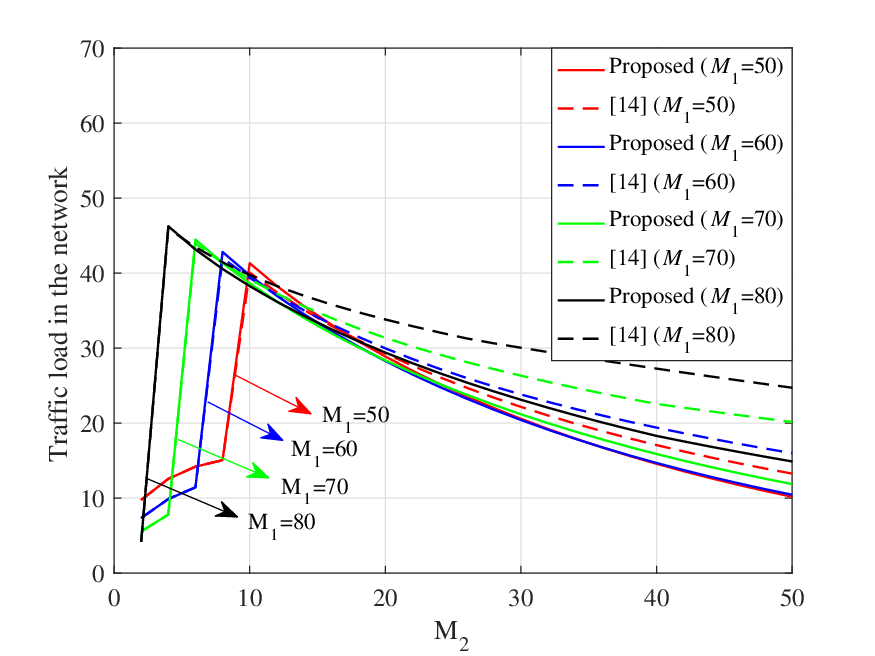}
%\centerline{\epsfig{file=Different_regions.eps,width=80mm}}
\caption{Comparison of the traffic load in the network with optimized $\alpha$ and $\beta$ in (14) and (15).}
\label{Optimized_traffic_load_comparison}
\end{figure}

Fig. \ref{Optimized_traffic_load_comparison} compares the traffic load in the network between the proposed caching scheme and the caching scheme in \cite{NKaram2016} as a function of $M_2$. In particular, we consider the network setting as follows: $F=1$ M bits, $N=100$, $K_1=10$, and $K_2=5$. In addition, the traffic load of the proposed algorithm is evaluated with the optimized $\alpha$ and $\beta$ in (14) and (15). The traffic load of the algorithm in \cite{NKaram2016} is evaluated with its optimized $\alpha$ and $\beta$. From the figure, we observe that, the proposed caching scheme achieves a traffic load similar to the caching scheme in \cite{NKaram2016} in Region I, i.e., $M_2$ is small and satisfies $M_1+K_2M_2<N$. Meanwhile, the proposed caching scheme is able to effectively reduce the traffic load compared with the caching scheme in \cite{NKaram2016} in Region II, i.e., $M_2$ is relatively large and satisfies $M_1+K_2M_2\geq N$. Furthermore, the traffic load gap of the two caching schemes becomes large as $M_2$ increases. This demonstrates the advantage of the proposed caching scheme compared with the caching scheme in \cite{NKaram2016}.

\section{Conclusions}

In this paper, we studied the decentralized caching in two-layer networks, in which users request contents from a server and the server satisfies users through helpers. By simultaneously utilizing the CCG, SMO, and CMO, we developed an hybrid caching scheme to reduce the traffic loads/transmission rates from both the server and the helpers. Besides, we analytically derived the transmission rates and demonstrated that the hybrid caching scheme is able to reduce the transmission rate from the server without increasing the transmission rate from each helper compared with the caching scheme in \cite{NKaram2016}. Furthermore, we theoretically analyzed the performance limit of the proposed caching scheme and demonstrated that the achievable rate region of the proposed caching scheme lies within constant margins to the information-theoretic optimum. In particular, the multiplicative and additive factors are carefully sharpened to be $\frac{1}{48}$ and $4$ respectively, both of which are better than those in \cite{NKaram2016}. Finally, simulation results demonstrated the advantage of the proposed caching scheme compared with the caching scheme in \cite{NKaram2016} in terms of the overall traffic load in the network.

%This reveals the advantages of the hybrid caching
%scheme over the generalized caching scheme in \cite{NKaram2016},
%where the constant multiplicative and additive factors are
%$\frac{1}{60}$ and $16$, respectively.

%\clearpage

\appendices

\section{Upper Bounds $R_1^{ub}(M_1,M_2)$ and $R_2^{ub}(M_2)$}

Since the transmission rates $(R_1^{H}(\alpha, \beta), R_2^{H}(\alpha,
\beta))$ are highly related to the values of $(\alpha, \beta)$, the
upper bounds $R_1^{ub}(M_1,M_2)$ and $R_2^{ub}(M_2)$ are also
determined by $(\alpha, \beta)$. Meanwhile, transmission
rates differ in variable $(M_1,M_2)$ regimes. Thus, we first
consider two regimes, i.e., Regime I) $M_1+K_2M_2\leq N$ and Regime
II) $M_1+K_2M_2> N$. In what follows, we will discuss the upper
bounds $R_1^{ub}(M_1,M_2)$ and $R_2^{ub}(M_2)$ in the two
regimes, respectively.

\subsection{Upper Bounds $R_1^{ub}(M_1,M_2)$ and $R_2^{ub}(M_2)$ in Regime I}
In this regime, we consider three tuples of $(\alpha, \beta)$. Tuple I is $(\alpha, \beta)=(\frac{M_1}{N}, \frac{M_1}{N})$, Tuple II is $(\alpha, \beta)=(\frac{M_1}{M_1+K_2M_2}, 0)$, Tuple III is $(\alpha, \beta)=(1, 1)$.
%\begin{align}
%(\alpha, \beta)=\left\{
%\begin{aligned}
%&  (\frac{M_1}{N}, \frac{M_1}{N}), \quad &  &\text{Tuple I}, \\
%&  (\frac{M_1}{M_1+K_2M_2}, 0), \quad &  &\text{Tuple II}, \\
%&  (1, 1), \quad &  &\text{Tuple III}, \\
%\end{aligned}
%\right.
%\label{Tuple_s_I}
%\end{align}

Firstly, we substitute $(\alpha, \beta)=(\frac{M_1}{N},
\frac{M_1}{N})$ into (\ref{Hybrid_Rate1}) and (\ref{Hybrid_Rate2}).
Then, we have
\begin{align} \nonumber
R_1^{H}(\alpha, \beta)=&\left(1-\frac{M_1}{N}\right)K_1K_2\left(1-\frac{M_2}{N}\right)\frac{N}{K_1K_2M_2}\times\\ \nonumber
&\left(1-\left(1-\frac{M_2}{N}\right)^{K_1K_2}\right)\leq \frac{N}{M_2}\left(1-\frac{M_2}{N}\right)\\
\leq & \min\left\{K_1K_2, \frac{N}{M_2}\left(1-\frac{M_2}{N}\right)\right\}
\label{R_1_ub_TI}
\end{align}
and
\begin{align}\nonumber
R_2^{H}(\alpha, \beta)=&K_2\left(1-\frac{M_2}{N}\right)\frac{N}{K_2M_2}\left(1-\left(1-\frac{M_2}{N}\right)^{K_2}\right)\\
\leq & \min\left\{K_2, \frac{N}{M_2}\right\}.
\label{R_2_ub_TI}
\end{align}

Substitute $(\alpha, \beta)=(\frac{M_1}{M_1+K_2M_2}, 0)$ into
(\ref{Hybrid_Rate1}) and (\ref{Hybrid_Rate2}), we have
\begin{align}\nonumber
R_1^{H}(\alpha, \beta)\leq & \frac{M_1K_2}{M_1\!+\!K_2M_2}\! \min\left\{K_1, \frac{N}{M_1\!+\!M_2K_2}\right\}\!+\\ \nonumber
&\!\frac{K_2M_2}{M_1\!+\!K_2M_2}\!\min\left\{K_1K_2,\frac{NK_2}{M_1\!+\!M_2K_2}\right\}\\ \nonumber
=& \frac{M_1}{M_1\!+\!K_2M_2}\! \min\left\{K_1K_2, \frac{NK_2}{M_1\!+\!M_2K_2}\right\}\!+\\ \nonumber
& \!\frac{K_2M_2}{M_1\!+\!K_2M_2}\!\min\left\{K_1K_2,\frac{NK_2}{M_1\!+\!M_2K_2}\right\}\\
\leq & \min\left\{K_1K_2, \frac{NK_2}{M_1+M_2K_2}\right\}
\label{R_1_ub_TII}
\end{align}
and
\begin{equation}
R_2^{H}(\alpha, \beta)\leq K_2 \overset{(a)}{=} \min\left\{K_2,
\frac{N}{M_2}\right\}, \label{R_2_ub_TII}
\end{equation}
where $(a)$ follows from $K_2M_2<N$ in regime I.

Substitute $(\alpha, \beta)=(1, 1)$ into (\ref{Hybrid_Rate1}) and
(\ref{Hybrid_Rate2}). Then, we have
\begin{align} \nonumber
R_1^{H}(\alpha, \beta)=&K_1K_2\left(1-\frac{M_1}{N}\right)\left(1-\frac{M_2}{N}\right)\frac{N}{K_1M_1}\times\\
&\left(\!1\!-\!\left(1\!-\!\frac{M_1}{N}\!\right)^{K_1}\!\right)=\frac{K_2N}{M_1}\left(\!1\!-\!\frac{M_1}{N}\!\right)
\label{R_1_ub_TIII}
\end{align}
and
\begin{equation}
\begin{split}
R_2^{H}(\alpha, \beta)\leq K_2 \overset{(a)}{=} \min\left\{K_2,
\frac{N}{M_2}\right\},
\end{split}
\label{R_2_ub_TIII}
\end{equation}
where $(a)$ follows from $K_2M_2<N$ in regime I.

Thus, if we choose $(\alpha^*, \beta^*)$ in three considered tuples corresponding to the
minimum $R_1^{H}(\alpha, \beta)$, we can
achieve the upper bounds $R_1^{ub}(M_1,M_2)$ and $R_2^{ub}(M_2)$
in Regime I as
\begin{align}\nonumber
R_1^{ub}(M_1,M_2)\leq & \min\left\{\!K_1K_2,\! \frac{N}{M_2}\left(1\!-\!\frac{M_2}{N}\right), \!\frac{NK_2}{M_1\!+\!M_2K_2}, \! \right.\\
&\left.\frac{K_2N}{M_1}\left(1\!-\!\frac{M_1}{N}\right)\right\}
\label{R_1_ub_I}.
\end{align}

and
 \begin{equation}
R_2^{ub}(M_2)\leq \min\left\{K_2, \frac{N}{M_2}\right\}.
\label{R_2_ub_I}
\end{equation}

\subsection{Upper Bounds $R_1^{ub}(M_1,M_2)$ and $R_2^{ub}(M_2)$ in Regime II}

In this regime, we consider the tuples of $(\alpha, \beta)$ as follows: Tuple I is $(\alpha, \beta)=(\frac{M_1}{N}, \frac{M_1}{N})$, and Tuple II is $(\alpha, \beta)=(\frac{M_1}{N}, \frac{1}{2})$.
%\begin{equation}
%(\alpha, \beta)=\left\{
%\begin{aligned}
%&  (\frac{M_1}{N}, \frac{M_1}{N}), \quad &  &\text{Tuple I}, \\
%&  (\frac{M_1}{N}, \frac{1}{2}), \quad &  &\text{Tuple II}. \\
%\end{aligned}
%\right.
%\label{Tuple_s_II}
%\end{equation}
Since we have obtained the upper bounds with Tuple I in (\ref{R_1_ub_TI}) and (\ref{R_2_ub_TI}), we will only calculate the upper bounds with Tuple II in the following.

Substitute $(\alpha, \beta)=(\frac{M_1}{N}, \frac{1}{2})$ into
(\ref{Hybrid_Rate1}) and (\ref{Hybrid_Rate2}), we have

\begin{align} \nonumber
R_1^{H}(\alpha, \beta)=&\left(1-\frac{M_1}{N}\right)\left(1-\frac{M_2}{2(N-M_1)}\right) \frac{2(N-M_1)}{M_2}\times\\ \nonumber
&\left(1-\left(1-\frac{M_2}{2(N-M_1)}\right)^{K_1K_2}\right)\\ \nonumber
\leq & \left(1-\frac{M_1}{N}\right) \min\left\{K_1K_2, \frac{2(N-M_1)}{M_2}\right\}\\
\leq & \min\left\{K_1K_2, \frac{2(N-M_1)^2}{NM_2}\right\}
\label{R_1_ub_TIII}
\end{align}
and
\begin{align} \nonumber
&R_2^{H}(\alpha, \beta)\\ \nonumber
=&\frac{M_1}{N}\left(1-\frac{M_2}{2M_1}\right)\frac{2M_1}{ M_2}\left(1-\left(1-\frac{M_2}{2M_1}\right)^{K_2}\right)\\ \nonumber
&+\left(1-\frac{M_1}{N}\right) \left(1-\frac{M_2}{2(N-M_1)}\right)\frac{2(N-M_1)}{M_2} \times\\ \nonumber
&\left(1-\left(1-\frac{M_2}{2(N-M_1)}\right)^{K_2}\right)\\ \nonumber
%\leq &\frac{M_1}{N} \min\left\{K_2,\frac{2M_1}{M_2}\right\}+\left(1-\frac{M_1}{N}\right) \min\left\{K_2,\frac{2(N-M_1)}{M_2}\right\}\\ \nonumber
\leq &\frac{M_1}{N}\min\left\{K_2,\frac{2N}{M_2}\right\}+\left(1-\frac{M_1}{N}\right) \min\left\{K_2,\frac{2N}{M_2}\right\}\\
=&\min\left\{K_2,\frac{2N}{M_2}\right\}
\leq 2\min\left\{K_2,\frac{N}{M_2}\right\}.
\label{R_2_ub_TIII}
\end{align}

Thus, if we choose $(\alpha^*, \beta^*)$ in two considered tuples corresponding to the
minimum $R_1^{H}(\alpha, \beta)$, we can
achieve the upper bounds $R_1^{ub}(M_1,M_2)$ and $R_2^{ub}(M_2)$
in Regime II as
 \begin{equation}
R_1^{ub}(M_1,\!M_2)\!\leq \!\min\left\{K_1K_2, \! \frac{N}{M_2}\left(1\!-\!\frac{M_2}{N}\right), \! \frac{2(N-M_1)^2}{NM_2}\right\}
\label{R_1_ub_II}
\end{equation}
and
 \begin{equation}
R_2^{ub}(M_2)\leq 2\min\left\{K_2, \frac{N}{M_2}\right\},
\label{R_2_ub_II}
\end{equation}
respectively.

\section{Gap between the Upper Bound $R_1^{ub}(M_1, M_2)$ and Lower Bound $R_1^{lb}(M_1, M_2)$}
In this section, we will characterize the gap between the upper bound $R_1^{ub}(M_1, M_2)$ and the lower bound $R_1^{lb}(M_1, M_2)$. Recall that we consider $K_1\geq 2$, $K_2\geq 2$, and $N\geq K_1K_2$.

\subsection{Gap between $R_1^{ub}(M_1, M_2)$ and $R_1^{lb}(M_1, M_2)$ in Regime I}

In regime I, we consider two subregimes, i.e., Subregime I) $0\leq M_1 \leq \frac{N}{2}$ and Subregime II) $\frac{N}{2} \leq M_1 \leq N$. Then, we will discuss the gap in the two subregimes respectively.

\subsubsection{Gap in Subregime I}
In this subregime, we have $M_1+K_2M_2\leq N$ and $0\leq M_1 \leq \frac{N}{2}$. Then, we have $0\leq M_2 \leq \frac{N-M_1}{K_2}\leq \frac{N}{2}$ using $K_2\geq 2$. Thus, we consider:

A) $0 \leq M_1< \frac{N}{2K_1}$ and $0 \leq M_2 \leq \frac{N}{K_1K_2}$;

B) $0 \leq M_1< \frac{N}{2K_1}$ and $\frac{N}{K_1K_2} \leq M_2 \leq \frac{N}{2K_2}$;

C) $0 \leq M_1< \frac{N}{2K_1}$ and $\frac{N}{2K_2} \leq M_2 \leq \frac{N}{2}$;

D) $\frac{N}{2K_1} \leq M_1< \frac{N}{4}$ and $0 \leq M_2 \leq \frac{N}{4K_2}$;

E) $\frac{N}{2K_1} \leq M_1< \frac{N}{4}$ and $\frac{N}{4K_2} \leq M_2 \leq \frac{N}{2}$;

F) $\frac{N}{4} \leq M_1< \frac{N}{2}$ and $0 \leq M_2 \leq \frac{N-M_1}{2K_2}$;

G) $\frac{N}{4} \leq M_1< \frac{N}{2}$ and $\frac{N-M_1}{2K_2} \leq M_2 \leq \frac{N-M_1}{K_2}$.

$\bullet$ A) $0 \leq M_1< \frac{N}{2K_1}$ and $0 \leq M_2 \leq \frac{N}{K_1K_2}$: we choose $s_1=\left \lfloor \frac{K_1}{2} \right \rfloor$ and $s_2=K_2$ in the lower bound (\ref{R_1_lb}). This is valid choice since $K_1\geq 2$, and thus $s_1=\left \lfloor \frac{K_1}{2} \right \rfloor \geq 1$. Then, we have
\begin{align} \nonumber
R_1^{lb}(M_1,M_2)&\geq \frac{\left \lfloor \frac{K_1}{2} \right \rfloor K_2 (N-\left \lfloor \frac{K_1}{2} \right \rfloor M_1-\left \lfloor \frac{K_1}{2} \right \rfloor K_2M_2)}{N+\left \lfloor \frac{K_1}{2} \right \rfloor K_2}\\  \nonumber &\overset{(a)}{\geq} \frac{\frac{K_1K_2}{4}(N-\frac{M_1K_1}{2}-\frac{M_2K_1K_2}{2})}{N+\frac{K_1K_2}{2}} \\ \nonumber
&\overset{(b)}{\geq} \frac{\frac{K_1K_2}{4}(N-\frac{N}{4}-\frac{N}{2})}{N+\frac{N}{2}}\\
&=\frac{K_1K_2}{24},
\end{align}
where $(a)$ follows from $x \geq \left \lfloor x \right \rfloor \geq \frac{x}{2}$ for any $x\geq 1$ and $(b)$ follows from $\frac{K_1M_1}{2}\leq \frac{N}{4}$ by using $M_1\leq \frac{N}{2K_1}$, $\frac{K_1K_2M_2}{2}\leq \frac{N}{2}$ by using $M_2\leq \frac{N}{K_1K_2}$, $\frac{K_1K_2}{2}\leq \frac{N}{2}$ by using $N \geq K_1K_2$. Combining with (\ref{R_1_ub_I}), we have
\begin{align} \nonumber
&R_1^{lb}(M_1,M_2) \geq \frac{K_1K_2}{24}
\geq \frac{1}{24} \min\left\{K_1K_2, \frac{N}{M_2}\left(1-\frac{M_2}{N}\right),\right.\\
& \left.\!\frac{NK_2}{M_1\!+\!M_2K_2}, \frac{K_2N}{M_1}\left(\!\!1\!-\!\frac{M_1}{N}\!\right)\!\right\}
\!\geq \! \frac{1}{24}R_1^{ub}(M_1,M_2).
\end{align}

$\bullet$ B) $0 \leq M_1< \frac{N}{2K_1}$ and $\frac{N}{K_1K_2} \leq M_2 \leq \frac{N}{2K_2}$: we choose $s_1=\left \lfloor \frac{N}{2M_2K_2} \right \rfloor$ and $s_2=K_2$ in the lower bound (\ref{R_1_lb}). Note that this is a valid choice since
\begin{equation}
1\overset{(a)}{\leq} \left \lfloor \frac{N}{2M_2K_2} \right \rfloor \overset{(b)}{\leq} \frac{N}{2M_2K_2} \overset{(c)}{\leq} \frac{K_1}{2},
\end{equation}
where $(a)$ follows from $M_2\leq \frac{N}{2K_2}$, $(b)$ follows from $x\geq\left \lfloor x \right \rfloor$ for any $x\geq 1$, and $(c)$ follows from $\frac{N}{K_1K_2}\leq M_2$.

Substitute $s_1$ and $s_2$ into the lower bound (\ref{R_1_lb}), we have
\begin{align} \nonumber
&R_1^{lb}(M_1,M_2)\\ \nonumber
\geq & \frac{\left \lfloor \frac{N}{2M_2K_2} \right \rfloor K_2 (N-\left \lfloor \frac{N}{2M_2K_2} \right \rfloor M_1-\left \lfloor \frac{N}{2M_2K_2} \right \rfloor K_2M_2)}{N+\left \lfloor \frac{N}{2M_2K_2} \right \rfloor K_2}\\ \nonumber
\overset{(a)}{\geq} & \frac{\frac{N}{4M_2}(N-\frac{NM_1}{2M_2K_2}-\frac{N}{2})}{N+\frac{N}{2M_2}}
= \frac{\frac{N}{4M_2}(\frac{1}{2}-\frac{M_1}{2M_2K_2})}{1+\frac{1}{2M_2}}\\
\overset{(b)}{\geq} & \frac{\frac{N}{4M_2}(\frac{1}{2}-\frac{1}{4})}{1+\frac{1}{2}}=\frac{N}{24M_2}\geq\frac{N}{24M_2}\left(1-\frac{M_2}{N}\right),
\end{align}
where $(a)$ follows from $x\geq\left \lfloor x \right \rfloor \geq \frac{x}{2}$ for any $x\geq 1$, $(b)$ follows from $\frac{M_1}{2M_2K_2} \leq \frac{N/(2K_1)}{2N/K_1} \leq \frac{1}{4}$ since $M_1 < \frac{N}{2K_1}$ and $M_2 \geq \frac{N}{K_1K_2}$, and $\frac{1}{2M_2} \leq \frac{K_1K_2}{2N} \leq \frac{1}{2}$ using $\frac{N}{K_1K_2}\leq M_2$ and $N \geq K_1K_2$. Combining with (\ref{R_1_ub_I}), we have
\begin{align} \nonumber
R_1^{lb}(M_1,M_2) \geq &\frac{N}{24M_2}\left(1-\frac{M_2}{N}\right)\\ \nonumber
\geq & \frac{1}{24} \min\left\{K_1K_2, \frac{N}{M_2}\left(1-\frac{M_2}{N}\right), \right.\\  \nonumber
& \left.\frac{NK_2}{M_1+M_2K_2}, \frac{K_2N}{M_1}\left(1-\frac{M_1}{N}\right)\right\} \\
\geq & \frac{1}{24}R_1^{ub}(M_1,M_2).
\end{align}

$\bullet$ C) $0 \leq M_1< \frac{N}{2K_1}$ and $\frac{N}{2K_2} \leq M_2 \leq \frac{N}{2}$: we choose $s_1=1$ and $s_2=\left \lfloor \frac{N}{2M_2} \right \rfloor$ in the lower bound (\ref{R_1_lb}). This is a valid choice since
\begin{equation}
1\leq \left \lfloor \frac{N}{2 \cdot N/2} \right \rfloor \overset{(a)}{\leq} \left \lfloor \frac{N}{2 M_2} \right \rfloor \overset{(b)}{\leq} \frac{N}{2 M_2} \overset{(c)}{\leq} K_2,
\label{C_valid_I}
\end{equation}
where $(a)$ follows from $M_2\leq \frac{N}{2}$, $(b)$ follows from $x\geq\left \lfloor x \right \rfloor$ for any $x\geq 1$, and $(c)$ follows from $\frac{N}{2K_2}\leq M_2$.

Substitute $s_1$ and $s_2$ into the lower bound (\ref{R_1_lb}), we have
\begin{align} \nonumber
&R_1^{lb}(M_1,M_2)\geq \frac{\left \lfloor \frac{N}{2 M_2} \right \rfloor(N-M_1-\left \lfloor \frac{N}{2 M_2} \right  \rfloor M_2)}{N+\left \lfloor \frac{N}{2M_2} \right \rfloor}\\ \nonumber
& \overset{(a)}{\geq} \frac{\frac{N}{4M_2}(N-M_1-\frac{N}{2})}{N+\frac{N}{2M_2}}= \frac{\frac{N}{4M_2}(\frac{1}{2}-\frac{M_1}{N})}{1+\frac{1}{2M_2}}\\
&\overset{(b)}{\geq} \frac{\frac{N}{4M_2}(\frac{1}{2}-\frac{1}{4})}{1+\frac{1}{2}}=\frac{N}{24M_2}\geq \frac{N}{24M_2}\left(1-\frac{M_2}{N}\right),
\end{align}
where $(a)$ follows from $x \geq \left \lfloor x \right \rfloor \geq \frac{x}{2}$ for any $x\geq 1$, $(b)$ follows from $\frac{M_1}{N}\leq \frac{N/(2K_1)}{N}\leq \frac{1}{4}$ using $M_1\leq \frac{N}{2K_1}$ and $K_1 \geq 2$, and $\frac{1}{2M_2} \leq \frac{K_2}{N}\leq\frac{K_1K_2}{K_1N} \leq \frac{1}{K_1}\leq \frac{1}{2}$ using $\frac{N}{2K_2}\leq M_2$, $N\geq K_1K_2$, and $K_1\geq 2$. Combining with (\ref{R_1_ub_I}), we have
\begin{align}
R_1^{lb}(M_1,M_2) \!\geq\!  \frac{N}{24M_2}\left(\!1\!-\!\frac{M_2}{N}\!\right)
\!\geq\!  \frac{1}{24}R_1^{ub}(M_1,M_2).
\end{align}

%\\ \nonumber
%\geq & \frac{1}{24} \min\left\{K_1K_2, \frac{N}{M_2}\left(1-\frac{M_2}{N}\right),\frac{NK_2}{M_1+M_2K_2}, \frac{K_2N}{M_1}\left(1-\frac{M_1}{N}\right)\right\}\\

%I-D) $0 \leq M_1< \frac{N}{2K_1}$ and $\frac{N}{4} \leq M_2 \leq N$: We trivially have
%\begin{equation}
%R_1^{lb} \geq 0 \geq \frac{N}{M_2}(1-\frac{M_2}{N})-3 \geq \frac{N}{M_2}(1-\frac{M_1}{N})(1-\frac{M_2}{N})-3 \geq \min\{K_1K_2, \frac{N}{M_2}(1-\frac{M_1}{N})(1-\frac{M_2}{N})\}-3.
%\end{equation}
%Combined with (\ref{R_1_ub_I}), we have
%\begin{equation}
%R_1^{lb} \geq R_1^{ub}-3.
%\end{equation}

$\bullet$ D) $\frac{N}{2K_1} \leq M_1< \frac{N}{4}$ and $0 \leq M_2 \leq \frac{N}{4K_2}$: we choose $s_1=\left \lfloor \frac{N}{2(M_1+M_2K_2)} \right \rfloor$ and $s_2=K_2$ in the lower bound (\ref{R_1_lb}). This is a valid choice since
\begin{align}\nonumber
&1=\left \lfloor \frac{N}{2( N/4+N/4)} \right \rfloor \overset{(a)}{\leq} \left \lfloor \frac{N}{2(M_1+M_2K_2)} \right \rfloor \\
 \overset{(b)}{\leq}& \frac{N}{2(M_1+M_2K_2)} \leq \frac{N}{2M_1} \overset{(c)}{\leq} K_1,
\end{align}
where $(a)$ follows from $M_1\leq \frac{N}{4}$ and $M_2\leq \frac{N}{4K_2}$, $(b)$ follows from $x\geq\left \lfloor x \right \rfloor$ for any $x\geq 1$, and $(c)$ follows from $\frac{N}{2K_1}\leq M_1$.

Then, we have
\begin{align} \nonumber
&R_1^{lb}(M_1,M_2)&\\ \nonumber
\geq & \frac{\left \lfloor \frac{N}{2(M_1+M_2K_2)} \right \rfloor K_2(N-\left \lfloor \frac{N}{2(M_1+M_2K_2)} \right \rfloor (M_1+M_2K_2))}{N+\left \lfloor \frac{N}{2(M_1+M_2K_2)} \right \rfloor K_2}\\ \nonumber
\overset{(a)}{\geq} &\frac{\frac{NK_2}{4(M_1+M_2K_2)}(N-\frac{N}{2(M_1+M_2K_2)}(M_1+K_2M_2))}{N+\frac{N K_2}{2(M_1+M_2K_2)}}\\ \nonumber
\overset{(b)}{\geq} &\frac{\frac{NK_2}{4(M_1+M_2K_2)} (N-\frac{N}{2})}{N+\frac{N K_2}{2M_1}}\\ \nonumber
\overset{(c)}{\geq} &\frac{\frac{NK_2}{4(M_1+M_2K_2)} (N-\frac{N}{2})}{N+N}\\
=&\frac{NK_2}{16(M_1+M_2K_2)},
\end{align}
where $(a)$ follows from $x \geq \left \lfloor x \right \rfloor \geq \frac{x}{2}$ for any $x\geq 1$, $(b)$ follows from $\frac{N K_2}{2(M_1+M_2k_2)} \leq \frac{N K_2}{2M_1}$, and $(c)$ follows from $\frac{NK_2}{2M_1} \leq K_1K_2 \leq N$ by using $\frac{N}{2K_1}\leq M_1$. Combining with (\ref{R_1_ub_I}), we have
\begin{align} \nonumber
R_1^{lb}(M_1,M_2) \geq & \frac{NK_2}{16(M_1+M_2K_2)}\\ \nonumber
\geq & \frac{1}{16} \min\left\{K_1K_2, \frac{N}{M_2}\left(1-\frac{M_2}{N}\right), \right.\\ \nonumber
& \left.\frac{NK_2}{M_1+M_2K_2}, \frac{K_2N}{M_1}\left(1-\frac{M_1}{N}\right)\right\}\\
\geq & \frac{1}{16} R_1^{ub}(M_1,M_2).
\label{E_lb}
\end{align}

$\bullet$ E) $\frac{N}{2K_1} \leq M_1< \frac{N}{4}$ and $\frac{N}{4K_2} \leq M_2 \leq \frac{N}{2}$: Let
\begin{equation}
(s_1, s_2)=\left\{
\begin{aligned}
& (\left \lfloor \frac{N}{4M_1} \right \rfloor, \left \lfloor \frac{M_1}{M_2} \right \rfloor), \quad & \text{if} \ M_1 \geq M_2, \\
& (\left \lfloor \frac{N}{4M_2} \right \rfloor, 1), \quad & \text{otherwise}.
\end{aligned}
\right.
\end{equation}
in the lower bound (\ref{R_1_lb}). This is a valid choice since for $M_1 \geq M_2$, we have
\begin{equation}
1=\left \lfloor \frac{N}{4 \cdot N/4} \right \rfloor \overset{(a)}{\leq} \left \lfloor \frac{N}{4M_1} \right \rfloor \overset{(b)}{\leq} \frac{N}{4M_1} \overset{(c)}{\leq} \frac{K_1}{2}
\end{equation}
and
\begin{equation}
1=\left \lfloor \frac{M_1}{M_2} \right \rfloor \leq  \frac{M_1}{M_2} \overset{(d)}{\leq} \frac{N/4}{N/(4K_2)}=K_2,
\end{equation}
where $(a)$ follows from $M_1 < \frac{N}{4}$, $(b)$ follows from $x \geq \left \lfloor x \right \rfloor \geq \frac{x}{2}$ for any $x\geq 1$, $(c)$ follows from $\frac{N}{2K_1} \leq M_1$, and $(d)$ follows from $M_1\leq \frac{N}{4}$ and $M_2\geq \frac{N}{4K_2}$.

For $M_1 < M_2$, we have
\begin{equation}
1=\left \lfloor \frac{N}{4 \cdot N/4} \right \rfloor \leq \left \lfloor \frac{N}{4M_2} \right \rfloor \leq \left \lfloor \frac{N}{4M_1} \right \rfloor \leq \frac{N}{4M_1} \leq \frac{K_1}{2}.
\end{equation}
Note that $s_1\leq \frac{N}{4M_1}$ and $\frac{N}{16M_2} \leq s_1s_2 \leq \frac{N}{4M_2}$ due to $x \geq \left \lfloor x \right \rfloor \geq \frac{x}{2}$ for any $x\geq 1$. Then, we have
\begin{align}\nonumber
&R_1^{lb}(M_1,M_2)\!\geq\! \frac{\frac{N}{16M_2} (N\!-\!\frac{N}{4M_1}M_1\!-\!\frac{N}{4M_2}M_2)}{N\!+\!\frac{N}{4M_2}}\!\\
&\overset{(a)}{\geq}\! \frac{\frac{N}{16M_2} (N\!-\!\frac{N}{2})}{N\!+\!\frac{N}{2}}\!=\!\frac{N}{48M_2}\!\geq\! \frac{N}{48M_2}\left(1\!-\!\frac{M_2}{N}\right),
\label{F_lb}
\end{align}
where $(a)$ follows from $\frac{N}{4M_2} \leq K_2 = \frac{K_1K_2}{K_1} \leq \frac{K_1K_2}{2} \leq \frac{N}{2}$ using $\frac{N}{4K_2}\leq M_2$, $N > K_1K_2$, and $K_1 \geq 2$. Combined with (\ref{R_1_ub_I}), we have
\begin{align}
R_1^{lb}(M_1,M_2)\!\geq \! \frac{N}{48M_2}\left(\!1\!-\!\frac{M_2}{N}\!\right)\! \geq\! \frac{1}{48} R_1^{ub}(M_1,M_2).
\end{align}
% \nonumber
%\geq & \frac{1}{48} \min\left\{K_1K_2, \frac{N}{M_2}\left(1-\frac{M_1}{N}\right), \frac{NK_2}{M_1+M_2K_2}, \frac{K_2N}{M_1}\left(1-\frac{M_1}{N}\right)\right\}\\

$\bullet$ F) $\frac{N}{4} \leq M_1< \frac{N}{2}$ and $0 \leq M_2 \leq \frac{N-M_1}{2K_2}$: we choose $s_1=1$ and $s_2=K_2$ in the lower bound (\ref{R_1_lb}). Then, we have
\begin{align} \nonumber
R_1^{lb}(M_1,M_2)&\geq \frac{ K_2(N-M_1-M_2K_2)}{N+K_2}\\ \nonumber
&\overset{(a)}{\geq} \frac{ K_2(N-M_1-\frac{N-M_2}{2})}{N+N/2}=\frac{K_2(N-M_1)}{3N}\\ \nonumber
&=\frac{K_2N}{3M_1}(1-\frac{M_1}{N})\cdot \frac{M_1}{N}
\overset{(b)}{\geq} \frac{K_2N}{3M_1}(1-\frac{M_1}{N})\cdot \frac{1}{4}\\
&= \frac{K_2N}{12M_1}(1-\frac{M_1}{N}),
\end{align}
where $(a)$ follows from $x \geq \left \lfloor x \right \rfloor \geq \frac{x}{2}$ for any $x\geq 1$ and $K_2\leq \frac{K_1K_2}{K_1}\leq \frac{N}{2}$ by using $N\geq K_1K_2$ and $K_1\geq 2$, $(b)$ follows from $\frac{M_1}{N}\geq \frac{N/4}{N}=\frac{1}{4}$. Combining with (\ref{R_1_ub_I}), we have
\begin{align} \nonumber
R_1^{lb}(M_1,M_2)\geq & \frac{K_2N}{12M_1}(1\!-\!\frac{M_1}{N})
\!\geq\! \frac{1}{12} \min\left\{K_1K_2,  \right.\\ \nonumber
&\left.\!\!\frac{N}{M_2}\left(1\!-\!\frac{M_2}{N}\right), \frac{NK_2}{M_1\!+\!M_2K_2}, \frac{K_2N}{M_1}(1\!-\!\frac{M_1}{N})\right\}\\
\geq &\frac{1}{12} R_1^{ub}(M_1,M_2).
\label{E_lb}
\end{align}

$\bullet$ G) $\frac{N}{4} \leq M_1< \frac{N}{2}$ and $\frac{N-M_1}{2K_2} \leq M_2 \leq \frac{N-M_1}{K_2}$: we choose $s_1=1$ and $s_2=\left \lfloor \frac{N-M_1}{2M_2} \right \rfloor$ in the lower bound (\ref{R_1_lb}). This is a valid choice since
\begin{equation}
1\overset{(a)}{\leq}\left \lfloor \frac{N-M_1}{K_2M_2} \right \rfloor \overset{(b)}{\leq} \left \lfloor \frac{N-M_1}{2M_2} \right \rfloor \overset{(c)}{\leq} \frac{N-M_1}{2M_2} \overset{(d)}{\leq} K_2,
\end{equation}
where $(a)$ follows from $M_1+K_2M_2 \leq N$, $(b)$ follows from $K_2\geq 2$, $(c)$ follows from $x \geq \left \lfloor x \right \rfloor$ for any $x\geq 1$, and $(d)$ follows from $\frac{N-M_1}{2K_2} \leq M_2$.

Then, we have
\begin{align} \nonumber
R_1^{lb}(M_1,M_2)&\!\geq \! \frac{\left \lfloor \frac{N\!-\!M_1}{2M_2} \right \rfloor(N\!-\!M_1\!-\!\left \lfloor \frac{N\!-\!M_1}{2M_2} \right \rfloor M_2)}{N\!+\!\left \lfloor \frac{N\!-\!M_1}{2M_2} \right \rfloor} \!\\  \nonumber
&\overset{(a)}{\geq} \! \frac{\frac{N\!-\!M_1}{4M_2}(N\!-\!M_1\!-\!\frac{N\!-\!M_1}{2M_2} M_2)}{N\!+\!N/2}\!=\!\frac{(N\!-\!M_1)^2}{12NM_2}\\ \nonumber
&=\frac{K_2(N\!-\!M_1)}{12M_1}\cdot \frac{N-M_1}{M_2K_2}\cdot \frac{M_1}{N}\!\\
&\overset{(b)}{\geq}\!\! \frac{K_2(N\!-\!M_1)}{12M_1}\cdot \!1\! \cdot \!\frac{1}{4}\!=\! \frac{K_2N}{48M_1}(1\!-\!\frac{M_1}{N}),
\end{align}
where $(a)$ follows from $x \geq \left \lfloor x \right \rfloor \geq \frac{x}{2}$ for any $x\geq 1$ and $\left \lfloor \frac{N-M_1}{2M_2} \right \rfloor \leq \frac{N-M_1}{2M_2} \leq K_2 \leq \frac{K_1K_2}{K_1}\leq \frac{N}{2}$ by using $\frac{N-M_1}{2K_2} \leq M_2$, $N\geq K_1K_2$, and $K_1\geq 2$, $(b)$ follows from $M_2 \leq \frac{N-M_1}{K_2}$ and $\frac{M_1}{N} \geq \frac{N/4}{N}=\frac{1}{4}$ by using $M_1\geq \frac{N}{4}$. Combining with (\ref{R_1_ub_I}), we have
\begin{align} \nonumber
&R_1^{lb}(M_1,M_2) \!\geq \!\frac{K_2N}{48M_1}\left(\!1\!-\!\frac{M_1}{N}\!\right)\\ \nonumber
&\!\geq \! \frac{1}{48} \min\left\{K_1K_2,\! \frac{N}{M_2}\left(\!1\!-\!\frac{M_2}{N}\!\right), \!\frac{NK_2}{M_1\!+\!M_2K_2}, \right.\\
& \left. \! \frac{K_2N}{M_1}\left(\!1\!-\!\frac{M_1}{N}\!\right)\right\}\geq \frac{1}{48} R_1^{ub}(M_1,M_2).
\label{E_lb}
\end{align}

\subsubsection{Gap between $R_1^{ub}(M_1,M_2)$ and $R_1^{lb}(M_1,M_2)$ in Subregime II}

In this subregime, i.e., $M_1+K_2M_2 \leq N$ and $\frac{N}{2}\leq M_1 \leq N$, we have $M_2\leq\frac{N-M_1}{K_2}$. Then, we consider two regions:

A) $\frac{N}{2} \leq M_1< N$ and $0 \leq M_2 \leq \frac{N-M_1}{2K_2}$;

B) $\frac{N}{2} \leq M_1< N$ and $\frac{N-M_1}{2K_2} \leq M_2 \leq \frac{N-M_1}{K_2}$,

$\bullet$ A) $\frac{N}{2} \leq M_1< N$ and $0 \leq M_2 \leq \frac{N-M_1}{2K_2}$: we choose $s_1=1$ and $s_2=K_2$ in the lower bound (\ref{R_1_lb}). Then, we have
\begin{align} \nonumber
&R_1^{lb}(M_1,M_2)\geq \frac{K_2 (N-M_1-K_2M_2)}{N+K_2}\\ \nonumber
&\overset{(a)}{\geq} \frac{K_2 (N\!-\!M_1\!-\!\frac{N-M_1}{2})}{N+N/2}\!=\!\frac{K_2(N\!-\!M_1)}{3N}= \frac{K_2N}{3N}\left(\!1\!-\!\frac{M_1}{N}\!\right)\\
&= \frac{K_2N}{6\cdot N/2}\left(1-\frac{M_1}{N}\right)\overset{(b)}{\geq} \frac{K_2N}{6M_1}\left(1-\frac{M_1}{N}\right),
\end{align}
where $(a)$ follows from $M_2 \leq \frac{N-M_1}{2K_2}$ and $K_2=\frac{K_1K_2}{K_1}\leq \frac{K_1K_2}{2} \leq \frac{N}{2}$ using $N>K_1K_2$ and $K_1 \geq 2$, and $(b)$ follows from $\frac{N}{2}\leq M_1$. Combining with (\ref{R_1_ub_I}), we have
\begin{align} \nonumber
&R_1^{lb}(M_1,M_2) \!\geq  \! \frac{K_2N}{6M_1}\left(\!1\!-\!\frac{M_1}{N}\!\right)\\ \nonumber
\geq & \!\frac{1}{6} \min\left\{\!\!K_1K_2, \! \frac{N}{M_2}\left(\!\!1\!-\!\frac{M_2}{N}\!\right),\!\frac{NK_2}{M_1\!+\!M_2K_2},\!\frac{K_2N}{M_1}\left(\!1\!-\!\frac{M_1}{N}\!\!\right)\!\!\right\}\\
\geq & \frac{1}{6} R_1^{ub}(M_1,M_2).
\end{align}

$\bullet$ B) $\frac{N}{2} \leq M_1< N$ and $\frac{N-M_1}{2K_2} \leq M_2 \leq \frac{N-M_1}{K_2}$: we choose $s_1=1$ and $s_2=\left \lfloor \frac{N-M_1}{2M_2} \right \rfloor$ in the lower bound (\ref{R_1_lb}). This is a valid choice since
\begin{equation}
1 \overset{(a)}{\leq} \left \lfloor \frac{N-M_1}{K_2M_2} \right \rfloor \overset{(b)}{\leq} \left \lfloor \frac{N-M_1}{2M_2} \right \rfloor \overset{(c)}{\leq} \frac{N-M_1}{2M_2} \overset{(d)}{\leq} K_2,
\end{equation}
where $(a)$ follows from $M_2 \leq \frac{N-M_1}{K_2}$, $(b)$ follows from $K_2\geq 2$, $(c)$ follows from $x\geq \left \lfloor x \right \rfloor$ for any $x\geq 1$, and $(d)$ follows from $\frac{N-M_1}{2K_2} \leq M_2$.

Substituting $s_1$ and $s_2$ in the lower bound (\ref{R_1_lb}), we obtain
\begin{align} \nonumber
&R_1^{lb}(M_1,M_2)\geq  \frac{\left \lfloor \frac{N-M_1}{2M_2} \right \rfloor(N-M_1-\left \lfloor \frac{N-M_1}{2M_2} \right \rfloor M_2)}{N+\left \lfloor \frac{N-M_1}{2M_2} \right \rfloor}\\ \nonumber
&\overset{(a)}{\geq} \frac{\frac{N-M_1}{4M_2}(N-M_1-\frac{N-M_1}{2M_2} M_2)}{N+\frac{N}{2}}= \frac{(N-M_1)^2}{12M_2N}\\
&\!= \!\frac{K_2N}{12M_1}\frac{N\!-\!M_1}{K_2M_2}\frac{M_1}{N}\left(\!1\!-\!\frac{M_1}{N}\!\right)\!\overset{(b)}{\geq} \! \frac{K_2N}{24M_1}\left(\!1\!-\!\frac{M_1}{N}\!\right),
\end{align}
where $(a)$ follows from $x\geq \left \lfloor x \right \rfloor \geq \frac{x}{2}$ for any $x\geq 1$, and
$\left \lfloor \frac{N-M_1}{2M_2} \right \rfloor \leq \frac{N-M_1}{2M_2} \leq K_2 \leq \frac{N}{K_1}\leq \frac{N}{2}$ using $\frac{N-M_1}{K_2} \leq M_2$, $N\geq K_1K_2$, and $K_1 \geq 2$, and $(b)$ follows from $\frac{N-M_1}{K_2M_2}\geq 1$ and $\frac{M_1}{N}\geq \frac{1}{2}$. Combining with (\ref{R_1_ub_I}), we have
\begin{align}
R_1^{lb}(M_1,M_2)\! \geq \!  \frac{K_2N}{24M_1}\left(\!1\!-\!\frac{M_1}{N}\!\right)\geq \frac{1}{24} R_1^{ub}(M_1,M_2).
\end{align}
%\\ \nonumber
%\geq &  \frac{1}{24} \min\left\{K_1K_2, \frac{N}{M_2}\left(1-\frac{M_2}{N}\right), \frac{NK_2}{M_1+M_2K_2}, \frac{K_2N}{M_1}(1-\frac{M_1}{N})\right\}\\

\subsection{Gap between $R_1^{ub}(M_1, M_2)$ and $R_1^{lb}(M_1, M_2)$ in Regime II}

In regime II, we consider two subregimes, i.e., Subregime I) $0\leq M_1 \leq \frac{N}{2}$ and Subregime II) $\frac{N}{2} \leq M_1 \leq N$. Then, we will discuss the gap in the two subregimes, respectively.

\subsubsection{Gap in Subregime I}
In this subregime, i.e., $M_1+K_2M_2\geq N$ and $0\leq M_1 \leq \frac{N}{2}$, we have $M_2\geq\frac{N-M_1}{K_2}\geq \frac{N}{2K_2} \geq \frac{N}{4K_2}$. Then, we consider:

A) $0 \leq M_1< \frac{N}{2K_1}$ and $\frac{N}{2K_2} \leq M_2 \leq \frac{N}{2}$;

B) $0 \leq M_1< \frac{N}{2K_1}$ and $\frac{N}{2} \leq M_2 \leq N$;

C) $\frac{N}{2K_1} \leq M_1< \frac{N}{4}$ and $\frac{N}{4K_2} \leq M_2 \leq \frac{N}{2}$;

D) $\frac{N}{2K_1} \leq M_1< \frac{N}{4}$ and $\frac{N}{2} \leq M_2 \leq N$;

E) $\frac{N}{4} \leq M_1< \frac{N}{2}$ and $\frac{N-M_1}{K_2} \leq M_2 \leq \frac{N-M_1}{2}$;

F) $\frac{N}{4} \leq M_1< \frac{N}{2}$ and $\frac{N-M_1}{2} \leq M_2 \leq N$.

$\bullet$ A) $0 \leq M_1< \frac{N}{2K_1}$ and $\frac{N}{2K_2} \leq M_2 \leq \frac{N}{2}$: we choose $s_1=1$ and $s_2=\left \lfloor \frac{N}{2M_2} \right \rfloor$ in the lower bound (\ref{R_1_lb}). This is a valid choice since
\begin{equation}
1\leq \left \lfloor \frac{N}{2 \cdot N/2} \right \rfloor \overset{(a)}{\leq} \left \lfloor \frac{N}{2 M_2} \right \rfloor \overset{(b)}{\leq} \frac{N}{2 M_2} \overset{(c)}{\leq} K_2,
\label{C_valid_I}
\end{equation}
where $(a)$ follows from $M_2\leq \frac{N}{2}$, $(b)$ follows from $x \geq \left \lfloor x \right \rfloor$ for any $x\geq 1$, and $(c)$ follows from $\frac{N}{2K_2} \leq M_2$.

Substitute $s_1$ and $s_2$ into the lower bound (\ref{R_1_lb}), we have
\begin{align} \nonumber
&R_1^{lb}(M_1,M_2)\geq \frac{\left \lfloor \frac{N}{2 M_2} \right \rfloor(N-M_1-\left \lfloor \frac{N}{2 M_2} \right \rfloor M_2)}{N+\left \lfloor \frac{N}{2M_2} \right \rfloor}\\ \nonumber
&\overset{(a)}{\geq} \frac{\frac{N}{4M_2}(N-M_1-\frac{N}{2})}{N+\frac{N}{2M_2}}= \frac{\frac{N}{4M_2}(\frac{1}{2}-\frac{M_1}{N})}{1+\frac{1}{2M_2}}\\
&\overset{(b)}{\geq} \frac{\frac{N}{4M_2}(\frac{1}{2}-\frac{1}{4})}{1+\frac{1}{2}}=\frac{N}{24M_2}\geq \frac{N}{24M_2}\left(1-\frac{M_2}{N}\right),
\end{align}
where $(a)$ follows from $x \geq \left \lfloor x \right \rfloor \geq \frac{x}{2}$ for any $x\geq 1$, $(b)$ follows from $\frac{M_1}{N}\leq \frac{N/(2K_1)}{N}=\frac{1}{2K_1} \leq \frac{1}{4}$ using $M_1\leq \frac{1}{2K_1}$ and $K_1\geq 2$, and $\frac{1}{2M_2} \leq \frac{K_2}{N}\leq\frac{K_1K_2}{K_1N} \leq \frac{1}{K_1}\leq \frac{1}{2}$ using $\frac{N}{2K_2} \leq M_2$, $N\geq K_1K_2$, and $K_1\geq 2$. Combining with (\ref{R_1_ub_I}), we have
\begin{align}
R_1^{lb}(M_1,M_2) \!\geq \!\frac{N}{24M_2}\left(\!1\!-\!\frac{M_2}{N}\!\right)\geq \frac{1}{24}R_1^{ub}(M_1,M_2).
\end{align}

%\\ \nonumber
%&\geq \frac{1}{24} \min\left\{K_1K_2, \frac{N}{M_2}\left(1-\frac{M_2}{N}\right), \frac{2(N-M_1)^2}{NM_2}\right\}\\
%&

$\bullet$ B) $0 \leq M_1< \frac{N}{2K_1}$ and $\frac{N}{2} \leq M_2 \leq N$: We have
\begin{equation}
R_1^{lb} \geq 0 \geq \frac{N}{M_2}-2 = \frac{N}{M_2}(1-\frac{M_2}{N})-1.
\end{equation}
Combining with (\ref{R_1_ub_II}), we have
\begin{align}
R_1^{lb}(M_1,M_2) &\geq \frac{N}{M_2}\left(1-\frac{M_2}{N}\right)-1\geq R_1^{ub}(M_1,M_2)-1.
\end{align}

%\\ \nonumber
%&\geq \min\left\{K_1K_2, \frac{N}{M_2}\left(1-\frac{M_2}{N}\right), \frac{2(N-M_1)^2}{NM_2}\right\}-1\\
%&

$\bullet$ C) $\frac{N}{2K_1} \leq M_1< \frac{N}{4}$ and $\frac{N}{4K_2} \leq M_2 \leq \frac{N}{2}$: Let
\begin{equation}
(s_1, s_2)=\left\{
\begin{aligned}
& (\left \lfloor \frac{N}{4M_1} \right \rfloor, \left \lfloor \frac{M_1}{M_2} \right \rfloor), \quad \text{if} \ M_1 \geq M_2, \\
& (\left \lfloor \frac{N}{4M_2} \right \rfloor, 1), \quad \quad \quad  \text{otherwise}.
\end{aligned}
\right.
\end{equation}
in the lower bound (\ref{R_1_lb}). This is a valid choice since for $M_1 \geq M_2$, we have
\begin{equation}
1=\left \lfloor \frac{N}{4 \cdot N/4} \right \rfloor \overset{(a)}{\leq} \left \lfloor \frac{N}{4M_1} \right \rfloor \overset{(b)}{\leq} \frac{N}{4M_1} \overset{(c)}{\leq} \frac{K_1}{2}
\end{equation}
and
\begin{equation}
1=\left \lfloor \frac{M_1}{M_2} \right \rfloor \leq  \frac{M_1}{M_2} \overset{(d)}{\leq} \frac{N/4}{N/(4K_2)}=K_2,
\end{equation}
where $(a)$ follows from $M_1\geq \frac{N}{4}$, $(b)$ follows from $x \geq \left \lfloor x \right \rfloor \geq \frac{x}{2}$ for any $x\geq 1$, $(c)$ follows from $M_1\geq \frac{N}{2K_1}$, $(d)$ follows from $M_1\leq \frac{N}{4}$ and $M_2 \geq \frac{N}{4K_2}$.

For $M_1 < M_2$, we have
\begin{equation}
1=\left \lfloor \frac{N}{4 \cdot N/4} \right \rfloor \leq \left \lfloor \frac{N}{4M_2} \right \rfloor \leq \left \lfloor \frac{N}{4M_1} \right \rfloor \leq \frac{N}{4M_1} \overset{(a)}{\leq} \frac{K_1}{2},
\end{equation}
where $(a)$ follows from $M_1\geq \frac{N}{2K_1}$.

Note that $s_1\leq \frac{N}{4M_1}$ and $\frac{N}{16M_2} \leq s_1s_2 \leq \frac{N}{4M_2}$ due to $x \geq \left \lfloor x \right \rfloor \geq \frac{x}{2}$ for any $x\geq 1$. Then, we have
\begin{align}\nonumber
&R_1^{lb}(M_1,M_2)\!\geq \!\frac{\frac{N}{16M_2} (N\!-\!\frac{N}{4M_1}M_1\!-\!\frac{N}{4M_2}M_2)}{N\!+\!\frac{N}{4M_2}}\\
&\overset{(a)}{\geq}\! \frac{\frac{N}{16M_2} (N\!-\!\frac{N}{2})}{N\!+\!\frac{N}{2}}\!=\!\frac{N}{48M_2}\!\geq\! \frac{N}{48M_2}\left(1\!-\!\frac{M_2}{N}\right),
\label{F_lb}
\end{align}
where $(a)$ follows from $\frac{N}{4M_2} \leq K_2 = \frac{K_1K_2}{K_1} \leq \frac{K_1K_2}{2} \leq \frac{N}{2}$ using $N > K_1K_2$ and $K_1 \geq 2$. Combined with (\ref{R_1_ub_I}), we have
\begin{align}
R_1^{lb}(M_1,M_2)&\geq \frac{N}{48M_2}\left(1-\frac{M_2}{N}\right)\geq \frac{1}{48} R_1^{ub}(M_1,M_2).
\end{align}

%\\ \nonumber
%&\geq \frac{1}{48} \min\left\{K_1K_2, \frac{N}{M_2}\left(1-\frac{M_2}{N}\right), \frac{2(N-M_1)^2}{NM_2}\right\}\\
%&

$\bullet$ D) $\frac{N}{2K_1} \leq M_1< \frac{N}{4}$ and $\frac{N}{2} \leq M_2 \leq N$: We have
\begin{equation}
R_1^{lb}(M_1,M_2) \geq 0 \geq \frac{N}{M_2}-2 \geq \frac{N}{M_2}(1-\frac{M_2}{N})-1.
\end{equation}
Combined with (\ref{R_1_ub_II}), we have
\begin{align}
R_1^{lb}(M_1,M_2)\!\geq \!\frac{N}{M_2}(1\!-\!\frac{M_2}{N})\!-\!1\geq R_1^{ub}(M_1,M_2)\!-\!1.
\end{align}
%\\ \nonumber
% &\geq \min\left\{K_1K_2, \frac{N}{M_2}\left(1-\frac{M_2}{N}\right), \frac{2(N-M_1)^2}{NM_2}\right\}-1\\
%&

$\bullet$ E) $\frac{N}{4} \leq M_1< \frac{N}{2}$ and $\frac{N-M_1}{K_2} \leq M_2 \leq \frac{N-M_1}{2}$: we choose $s_1=1$ and $s_2=\left \lfloor \frac{N-M_1}{2M_2} \right \rfloor$ in the lower bound (\ref{R_1_lb}). This is a valid choice since
\begin{equation}
1\leq  \left \lfloor \frac{N-M_1}{2 M_2} \right \rfloor \leq \frac{N-M_1}{2 M_2} \overset{(a)}{\leq} \frac{K_2}{2},
\label{C_valid_I}
\end{equation}
where $(a)$ follows from $\frac{N-M_1}{K_2} \leq M_2$.

Substitute $s_1$ and $s_2$ into the lower bound (\ref{R_1_lb}), we have
\begin{align}\nonumber
R_1^{lb}(M_1,M_2)\!\geq &\!\frac{ \left \lfloor \frac{N\!-\!M_1}{2 M_2} \right \rfloor(N\!-\!M_1\!-\! \left \lfloor \frac{N\!-\!M_1}{2 M_2} \right \rfloor M_2)}{N\!+\! \left \lfloor \frac{N\!-\!M_1}{2 M_2} \right \rfloor}\\ \nonumber
\overset{(a)}{\geq} & \!\frac{\frac{N\!-\!M_1}{4 M_2}(N\!-\!M_1\!-\!\frac{N\!-\!M_1}{2})}{N\!+\!\frac{N}{4}}\\
=& \frac{(N\!-\!M_1)^2}{10NM_2},
\end{align}
where $(a)$ follows from $x \geq \left \lfloor x \right \rfloor \geq \frac{x}{2}$ for any $x\geq 1$, and $\left \lfloor \frac{N-M_1}{2 M_2} \right \rfloor \leq \left \lfloor \frac{K_2}{2} \right \rfloor \leq \frac{K_2}{2}\leq \frac{K_1K_2}{2K_1}\leq \frac{N}{4}$ using $\frac{N-M_1}{K_2} \leq M_2$, $K_1K_2\leq N$, and $K_1\geq 2$. Combining with (\ref{R_1_ub_I}), we have
\begin{align}\nonumber
&R_1^{lb}(M_1,\!M_2)\geq \frac{(N\!-\!M_1)^2}{10NM_2} \!\geq \!\frac{1}{20} \min\!\left\{K_1K_2, \frac{N}{M_2}\left(\!1\!-\!\frac{M_2}{N}\right),\right.\\
&\left. \frac{2(N-M_1)^2}{NM_2}\right\}\geq \frac{1}{20}R_1^{ub}(M_1,\!M_2).
\end{align}

$\bullet$ F) $\frac{N}{4} \leq M_1< \frac{N}{2}$ and $\frac{N-M_1}{2} \leq M_2 \leq N$: We have
\begin{align} \nonumber
&R_1^{lb}(M_1,M_2)\geq 0= \frac{2(N-M_1)^2}{NM_2}-\frac{2(N-M_1)^2}{NM_2}\\ \nonumber
&= \frac{2(N-M_1)^2}{NM_2}-\frac{N-M_1}{M_2}\cdot \frac{2(N-M_1)}{N}\\
&\overset{(a)}{\geq} \frac{2(N-M_1)^2}{NM_2}-2\cdot 2= \frac{2(N-M_1)^2}{NM_2}-4,
\end{align}
where $(a)$ follows from $\frac{N-M_1}{2} \leq M_2$ and $\frac{N-M_1}{N}\leq 1$.

Combined with (\ref{R_1_ub_II}), we have
\begin{align} \nonumber
&R_1^{lb}(M_1,M_2)\geq \frac{2(N-M_1)^2}{NM_2}-4\geq \min\left\{K_1K_2, \right.\\
&\left.\!\frac{N}{M_2}\left(\!1\!-\!\frac{M_2}{N}\!\right), \!\frac{2(N\!-\!M_1)^2}{NM_2}\!\!\right\}\!-\!4\!\geq \! R_1^{ub}(M_1,M_2)\!-\!4.
\end{align}

\subsubsection{Gap in Subregime II}
In this subregime, i.e., $M_1+K_2M_2\geq N$ and $\frac{N}{2}\leq M_1 \leq N$, we have $M_2 \geq \frac{N-M_1}{K_2}$. Then, we consider:

 A) $\frac{N}{2} \leq M_1< N$ and $\frac{N-M_1}{K_2} \leq M_2 \leq \frac{N-M_1}{2}$;

 B) $\frac{N}{2} \leq M_1< N$ and $\frac{N-M_1}{2} \leq M_2 \leq N$.

$\bullet$ A) $\frac{N}{2} \leq M_1< N$ and $\frac{N-M_1}{K_2} \leq M_2 \leq \frac{N-M_1}{2}$: we choose $s_1=1$ and $s_2=\left \lfloor \frac{N-M_1}{2M_2} \right \rfloor$ in the lower bound (\ref{R_1_lb}). This is a valid choice since
\begin{equation}
1 \leq \left \lfloor \frac{N-M_1}{2M_2} \right \rfloor \overset{(a)}{\leq} \frac{N-M_1}{2M_2} \overset{(b)}{\leq} \frac{K_2}{2},
\end{equation}
where $(a)$ follows from $x \geq \left \lfloor x \right \rfloor $ for any $x\geq 1$, $(b)$ follows from $\frac{N-M_1}{K_2} \leq M_2$.

Then, we obtain
\begin{align}\nonumber
R_1^{lb}(M_1,M_2)\geq &  \frac{\left \lfloor \frac{N\!-\!M_1}{2M_2} \right \rfloor(N\!-\!M_1\!-\!\left \lfloor \frac{N\!-\!M_1}{2M_2} \right \rfloor M_2)}{N\!+\!\left \lfloor \frac{N\!-\!M_1}{2M_2} \right \rfloor}\\ \nonumber
\overset{(a)}{\geq} & \frac{\frac{N\!-\!M_1}{4M_2}(N\!-\!M_1\!-\!\frac{N\!-\!M_1}{2M_2} M_2)}{N\!+\!\frac{N}{4}}\\
= & \frac{(N\!-\!M_1)^2}{10M_2N},
\end{align}
where $(a)$ follows from $x \geq \left \lfloor x \right \rfloor \geq \frac{x}{2}$ for any $x\geq 1$ and $\left \lfloor \frac{N-M_1}{2M_2} \right \rfloor \leq \left \lfloor \frac{K_2}{2} \right \rfloor \leq \frac{K_2}{2} \leq \frac{K_1K_2}{2K_1}\leq \frac{N}{4}$
using $\frac{N-M_1}{K_2} \leq M_2$, $N\geq K_1K_2$, and $K_1 \geq 2$. Combining with (\ref{R_1_ub_II}), we have
\begin{align}
R_1^{lb} (M_1,M_2)&\geq \frac{(N-M_1)^2}{10M_2N}\geq\frac{1}{20}R_1^{ub}(M_1,M_2).
\end{align}

%\\ \nonumber
% &\geq \frac{1}{20}\min\left\{K_1K_2, \frac{N}{M_2}\left(1-\frac{M_2}{N}\right), \frac{2(N-M_1)^2}{NM_2}\right\}\\
%&

$\bullet$ B) $\frac{N}{2} \leq M_1< N$ and $\frac{N-M_1}{2} \leq M_2 \leq N$: We have
\begin{align} \nonumber
&R_1^{lb}(M_1,M_2) \geq 0=\frac{2(N-M_1)^2}{NM_2}-\frac{2(N-M_1)^2}{NM_2}\\ \nonumber
&= \frac{2(N-M_1)^2}{NM_2}-\frac{N-M_1}{M_2}\cdot \frac{2(N-M_1)}{N}\\
&\overset{(a)}{\geq} \frac{2(N-M_1)^2}{NM_2}-2\cdot 2= \frac{2(N-M_1)^2}{M_2N}-4,
\end{align}
where $(a)$ follows from $\frac{N-M_1}{2} \leq M_2$ and $\frac{N-M_1}{N} \leq 1$.

Combining with (\ref{R_1_ub_II}), we have
\begin{align}
R_1^{lb}M_1,M_2) \!\geq\! \frac{2(N-M_1)^2}{M_2N}\!-\!4\!\geq\! R_1^{ub}(M_1,M_2)\!-\!4.
\end{align}
%\\ \nonumber
%& \geq \min\left\{K_1K_2, \frac{N}{M_2}\left(1-\frac{M_2}{N}\right), \frac{2(N-M_1)^2}{NM_2}\right\}-4\\
%&

Combining the results in Subsections A and B, we have that the upper bound $R_1^{ub}$ and the lower bound $R_1^{lb}$ are within a constant multiplicative and additive gap for all pairs of $M_1$ and $M_2$. More specifically, we have
\begin{equation}
R_1^{lb}(M_1,M_2) \geq \frac{1}{48}R_1^{ub}(M_1,M_2)-4.
\label{R_2_result_B}
\end{equation}

\section{Gap between the Upper Bound $R_2^{ub}(M_2)$ and Lower Bound $R_2^{lb}(M_2)$}
In this section, we will characterize the gap between the upper
bound $R_2^{ub}(M_2)$ and the lower bound $R_2^{lb}(M_2)$.
We also consider $K_1\geq 2$, $K_1\geq 2$, and $N\geq K_1K_2$.
Recall that lower bound and upper bound of the transmission rate
$R_2(\alpha^*,\beta^*)$ are
\begin{equation}
R_2^{lb}(M_2)\triangleq \underset{t \in [K_2]}{\max} \frac{t(N-tM_2)}{N+t}
\label{R_2_lb}
\end{equation}
and
\begin{equation}
R_2^{ub}(M_2)=2\min\left\{K_2, \frac{N}{M_2}\right\},
\label{R_2_ub}
\end{equation}
respectively.

To discuss the gap between the lower and the upper bound, we consider two regimes:

A) $0\leq M_2 < \frac{N}{2} $;

B) $\frac{N}{2} \leq M_2 \leq N$.

$\bullet$ A) $0\leq M_2 < \frac{N}{2} $: we choose $t=\left \lfloor
\frac{1}{2} \min\{K_2, \frac{N}{M_2}\} \right \rfloor$ in the lower
bound (\ref{R_2_lb}). This is a valid choice since
\begin{equation}
1=\left \lfloor \frac{1}{2} \min\{K_2, \frac{N}{M_2}\} \right \rfloor \leq \frac{K_2}{2}.
\end{equation}
Then, we have
\begin{align} \nonumber
R_2^{lb}(M_2)\geq & \frac{\left \lfloor \frac{1}{2} \min\{K_2, \frac{N}{M_2}\} \right \rfloor \left(N-\left \lfloor \frac{1}{2} \min\{K_2, \frac{N}{M_2}\} \right \rfloor M_2\right)}{N+\left \lfloor \frac{1}{2} \min\{K_2, \frac{N}{M_2}\} \right \rfloor}\\ \nonumber
\overset{(a)}{\geq} &\frac{\frac{1}{4} \min\{K_2, \frac{N}{M_2}\} \left(N-\frac{N}{2}\right)}{N+\frac{N}{4}}\\
=&\frac{1}{10} \min\{K_2, \frac{N}{M_2}\},
\label{R_2_lb_A}
\end{align}
where $(a)$ follows from $x \geq \left \lfloor x \right \rfloor \geq \frac{x}{2}$ for any $x\geq 1$ and
$\left \lfloor \frac{1}{2} \min\{K_2, \frac{N}{M_2}\} \right \rfloor \leq \frac{K_2}{2} \leq \frac{K_1K_2}{2K_1} \leq \frac{N}{4}$
using $N\geq K_1K_2$ and $K_1 \geq 2$. Combining with (\ref{R_2_ub}), we have
\begin{align}\nonumber
&R_2^{lb}(M_2) \geq \frac{1}{10} \min\{K_2, \frac{N}{M_2}\}\geq \frac{1}{20} \cdot 2\min\{K_2, \frac{N}{M_2}\}\\
&\geq \frac{1}{20}R_2^{ub}(M_2).
\label{R_2_result_A}
\end{align}

$\bullet$ B) $\frac{N}{2} \leq M_2 < N $: We have
\begin{equation}
R_2^{lb}(M_2) \geq 0=2\frac{N}{M_2}-2\frac{N}{M_2}.
\end{equation}
Combining with (\ref{R_2_ub}), we have
\begin{align}\nonumber
&R_2^{lb}(M_2) \geq 2\frac{N}{M_2}-2\frac{N}{M_2}\geq 2\min\{K_2, \frac{N}{M_2}\}-4\\
&\geq R_2^{ub}(M_2)-4.
\label{R_2_result_B}
\end{align}

By combining (\ref{R_2_result_A}) and (\ref{R_2_result_B}), we have
\begin{equation}
R_2^{lb}(M_2) \geq \frac{1}{20}R_2^{ub}(M_2)-4.
\end{equation}

\bibliographystyle{IEEEtran}
\bibliography{IEEEabrv,DecentralRefZhao}

\end{document}